\documentclass[11pt]{article}
\usepackage[letterpaper,margin=1in]{geometry}
\IfFileExists{lmodern.sty}{\usepackage[T1]{fontenc}\usepackage{lmodern}}{}
\usepackage[expansion=false]{microtype}
\usepackage{setspace}
\usepackage{amsmath,amssymb,amsthm,bm}
\usepackage{enumitem,etoolbox,needspace,booktabs,threeparttable,array}\usepackage{graphicx}\graphicspath{{figures/}}\usepackage[font=small,labelfont=bf,labelsep=period,justification=justified,singlelinecheck=false]{caption}\newcommand{\fignote}[1]{\par\vspace{3pt}\begin{minipage}{\linewidth}\footnotesize\emph{Notes:} #1\end{minipage}}

\usepackage[authoryear,round]{natbib}
\usepackage[dvipsnames]{xcolor}
\usepackage[colorlinks=true,allcolors=MidnightBlue,pdfusetitle]{hyperref}
\newtheorem{assumption}{Assumption}
\newtheorem{definition}{Definition}
\newtheorem{lemma}{Lemma}
\newtheorem{proposition}{Proposition}
\newtheorem{theorem}{Theorem}
\newtheorem{corollary}{Corollary}
\theoremstyle{remark}
\newtheorem{remark}{Remark}
\makeatletter
\renewenvironment{proof}[1][\proofname]{\par
  \pushQED{\qed}\normalfont\topsep6\p@\@plus6\p@\relax
  \trivlist\item[\hskip\labelsep\itshape #1\@addpunct{.}]
  \itshape\ignorespaces
}{\popQED\endtrivlist\@endpefalse}
\makeatother
\BeforeBeginEnvironment{assumption}{\Needspace{7\baselineskip}}
\DeclareMathOperator{\rank}{rank}
\DeclareMathOperator{\col}{col}
\DeclareMathOperator{\diag}{diag}
\DeclareMathOperator{\argmin}{arg\,min}
\DeclareMathOperator{\Var}{Var}
\DeclareMathOperator{\Cov}{Cov}
\newcommand{\R}{\mathbb R}
\newcommand{\E}{\mathbb E}
\newcommand{\Prob}{\mathbb P}
\newcommand{\one}{\bm 1}
\newcommand{\F}{\mathcal F_{\mathrm{tr}}}
\newcommand{\Pone}{\mathcal P_1}
\newcommand{\Ptwo}{\mathcal P_2}
\newcommand{\Q}{\mathcal Q}
\newcommand{\K}{\mathcal K}
\newcommand{\cv}{\operatorname{cv}_{\alpha}}
\newcommand{\norm}[1]{\lVert#1\rVert}
\newcommand{\abs}[1]{\lvert#1\rvert}
\newcommand{\pos}[1]{(#1)_{+}}
\title{\textbf{Honest Inference for Synthetic Controls\linebreak with Imperfect Fit}\vspace{3pt}\\\large Time-Split Validity and the Choice of Anchor}
\author{Rok Spruk\thanks{Associate Professor of Economics, School of Economics and Business, University of Ljubljana; University of Western Australia. Email: rok.spruk@ef.uni-lj.si}}
\date{}
\hypersetup{pdftitle={Honest Inference for Synthetic Controls with Imperfect Fit},pdfauthor={}}
\begin{document}
\maketitle
\vspace{-2em}
\begin{abstract}
Synthetic controls are used for one treated unit that lies outside its donors’ range, with a pre-treatment window too short for residual-based standard errors to be trusted. We propose inference that separates the periods used to fit the synthetic control from the periods used to assess its error. Weights are fitted on an early block of pre-treatment periods, and the gap on a later block calibrates the bias and the noise. The resulting interval has finite-sample coverage under an explicit restriction on how far the counterfactual may depart from the extrapolated pre-treatment gap, for any fitting rule that uses only the early block. The leading restriction bounds the curvature of the gap and leaves its slope free, and the breakdown curvature at which the interval first includes zero states how much extrapolation a conclusion needs. We show when drift bounds calibrated from donor placebos are valid and how to allow for serial dependence. In simulations, conventional intervals undercover when the treated unit drifts from its donors, while the proposed interval covers whenever its restriction holds and the noise is serially independent. Re-analyses of three classic applications show that some well-known conclusions rest on far stronger extrapolation than others.
\end{abstract}

\section{Introduction}\label{sec:introduction}
Synthetic controls estimate the effect of an intervention on one unit by comparing it with a weighted combination of untreated units \citep{abadie2003,abadie2010}. In a short panel the pre-treatment outcomes must select the weights, assess the fit and inform the uncertainty, and these tasks place conflicting demands on the same observations. When the treated unit lies outside the donors' range, exact fit is infeasible and imperfect fit can carry bias that does not vanish with more periods \citep{ferman2021}. The synthetic unit then drifts away from the treated unit as soon as it is evaluated on periods it was not fitted to, and the question is how far that drift may be extrapolated into the post-treatment period.

This paper answers with a time split and an explicit sensitivity class. Donor weights and a temporal anchor are chosen on a training block of pre-treatment periods. The estimator subtracts the anchored gap on a later validation block from the average post-treatment gap. Conditional on training, its error is a bounded bias plus a Gaussian disturbance, and an orthogonal projection of the validation residuals gives a scale estimate independent of that disturbance. Adding the worst-case bias to a Student critical radius then gives an interval with coverage at least its nominal level in finite samples, uniformly over the class, for any fitting rule that uses the training block, including the weights of \citet{abadie2010}, \citet{doudchenko2016}, \citet{benmichael2021} and \citet{hollingsworth2022}, and with as few as three validation periods, although with so few the Student critical value is too large to be useful.

The class that matters in practice bounds the curvature of the untreated discrepancy and leaves its slope free. Under it the interval is valid whatever the drift, and its length is governed by one number, the bound on how fast the slope may change. The breakdown curvature, at which the interval first includes zero, is the summary we propose. It states in the units of the outcome how much bending of the counterfactual path a conclusion can tolerate, and comparing it with the curvature that donor placebos display tells the reader whether the conclusion rests on a restriction the donors satisfy. A linear class, in which the slope may be bounded, is the special case with no curvature, and the anchor that minimises the length of the interval under it is characterised in closed form.

Bounds have to come from somewhere, and the donors are the obvious source. A bound computed from placebo errors is not a function of the training block, so the conditional argument does not cover it. We show that a placebo-calibrated drift bound, inflated by the treated unit's own noise and used with a critical value that is simultaneous over a family of anchors, covers with probability at least $1-\alpha-1/(J+1)$ when the treated unit's placebo error is exchangeable with the donors', and that the anchor may then be chosen after the data are seen. The term $1/(J+1)$ bounds the price of calibrating from $J$ donors, and the assumption is the one on which placebo inference rests \citep{abadie2010,firpo2018}. Serial dependence, the other assumption a short panel cannot check, is handled by two calibrations valid under bounded dependence and by a breakdown persistence.

The evidence has two parts. A simulation that re-estimates the weights, calibrates the bounds and selects the anchor in every replication shows that the conventional construction, which fits weights and estimates noise on the same periods, covers 82 percent of the time at a nominal 95 even when the treated unit's growth is drawn like the donors', that the time-split interval covers whenever its bounds hold and the disturbances are independent, that the placebo-calibrated bound fails exactly when the treated unit drifts more than every donor, and that the prediction intervals of \citet{cattaneo2021} are the strongest competitor, while the $t$-test of \citet{chernozhukov2021t} undercovers in panels this short. The three applications on which the method was introduced, California's Proposition 99 \citep{abadie2010}, German reunification \citep{abadie2015} and Basque terrorism \citep{abadie2003}, show the breakdown curvature at work. The reunification result survives more curvature than any placebo displays, though less once 1990 is excluded from the validation block, the Basque result survives the curvature of most placebos, and the California result survives only a restriction that most placebos violate.

The rest of the paper is organized as follows. Section~\ref{sec:literature} relates the analysis to the literature. Section~\ref{sec:theory} presents the model, the estimator, the curvature class and the placebo calibration. Section~\ref{sec:dependence} gives the calibrations under bounded dependence. Section~\ref{sec:evidence} reports the simulations and Section~\ref{sec:application} the applications. Section~\ref{sec:discussion} discusses the scope of the guarantees and Section~\ref{sec:conclusion} concludes. Proofs are in Appendix~\ref{app:proofs}.

\section{Related Literature}\label{sec:literature}

The paper sits at the junction of four literatures, on synthetic-control estimation, on inference for a single treated unit, on sensitivity analysis under bounded departures from an identifying restriction, and on sample splitting for valid inference after data-dependent selection. We take them in turn, and we state at the end what the paper adds to each.

\subsection{Synthetic-Control Estimation and Imperfect Fit}

In the seminal contribution, \citet{abadie2003} and \citet{abadie2010} introduced the synthetic-control method as a transparent way to build a comparison unit for a single treated aggregate from a convex combination of untreated donors, with weights chosen to reproduce the treated unit's pre-treatment outcomes and predictors. In future work, \citet{abadie2015} extended the method to comparative politics, and, more recently, \citet{abadie2021} surveys the method's feasibility conditions, data requirements and the role of the convex-hull restriction, noting that the method is ill suited to a treated unit whose characteristics lie far outside the donors' range. The convex-hull restriction is what the present paper relaxes, and the resulting level discrepancy is what the intercept in our discrepancy class absorbs.

The early literature treated perfect pre-treatment fit as a prerequisite. \citet{ferman2021} show that imperfect fit does not merely add noise, since when assignment is correlated with latent confounders the bias of the synthetic-control estimator persists even as the number of pre-treatment periods grows, and they study demeaning as a partial remedy. For instance, \citet{ferman2021b} characterises the asymptotic behaviour of the estimator with many periods and many donors, and \citet{ferman2020} show how the freedom to choose predictors and specifications can be used to select results, which motivates the pre-specification of the split and the sensitivity constants that we require. \citet{botosaru2019} examine the role of covariates in the weighting problem, and \citet{kaul2022} show that using every pre-treatment outcome as a predictor renders covariates irrelevant to the weights. \citet{klossner2018} document the sensitivity of a well-known application to the optimisation procedure, which is one more reason to separate the fitting stage from the stage at which inference is conducted.

A second strand modifies the estimator to handle imperfect fit directly. \citet{doudchenko2016} unify synthetic control, difference-in-differences and regression-based imputation in a single framework that allows an intercept, unrestricted or penalised weights and weights that need not sum to one, and they are the origin of the de-meaned estimator we use. \citet{benmichael2021} augment the synthetic-control estimate with an outcome model that corrects the residual imbalance, which permits controlled extrapolation and negative donor weights, and \citet{benmichael2022} extend the idea to staggered adoption. \citet{arkhangelsky2021} combine regularised unit weights with time weights, so that the counterfactual is anchored on a weighted average of the pre-treatment periods chosen to predict the donors' post-treatment outcomes, which reduces the bias from latent factors that move over time. Their time weights are an anchor in the sense of this paper, chosen to predict the donors' post-treatment outcomes, and Theorem~\ref{thm:affine} is an alternative criterion for such weights that minimises the length of a bias-aware interval instead. \citet{abadie2021b} penalise the weights toward nearest neighbours for disaggregated data, and \citet{kellogg2021} trade off the interpolation bias of matching against the extrapolation bias of synthetic control. \citet{hollingsworth2022} fit the weights by lasso with an intercept and unrestricted signs, chosen by cross-validation that respects the order of time. All of these estimators are anchored in the sense of Section~\ref{sec:estimator}, since each subtracts a weighted average of the treated unit's own pre-treatment gaps, and their unit weights are covered by the inference result of Section~\ref{sec:split} when they are fitted on the training block. Time weights chosen, as in \citet{arkhangelsky2021}, to predict the donors' post-treatment outcomes are not, since they depend on donor outcomes outside the training block.

A third strand replaces the weighting problem by an explicit model of untreated outcomes. \citet{bai2009} studies panel models with interactive fixed effects, \citet{gobillon2016} compare them with synthetic controls for regional policy evaluation, and \citet{xu2017} builds the generalised synthetic-control estimator on them. \citet{athey2021} impute the missing counterfactual by nuclear-norm matrix completion, \citet{amjad2018} use a singular-value threshold to denoise the donor matrix before fitting weights, and \citet{masini2021} and \citet{carvalho2018} build artificial counterfactuals from high-dimensional time-series regressions with inference that allows nonstationarity. \citet{chen2023} interprets synthetic control as online linear regression and derives regret bounds, and \citet{gunsilius2023} extends the method to distributional outcomes. \citet{arkhangelsky2024} survey the field. In short panels the factor loadings that these models estimate for the treated unit are weakly identified, and our discrepancy class is designed to absorb the imbalance they leave rather than to estimate it.

\subsection{Inference with a Single Treated Unit}

Inference for synthetic control has rested mainly on permutation. By way of example, \citet{abadie2010} rank the treated unit's post-treatment to pre-treatment prediction error among in-space placebos, and \citet{firpo2018} formalise the exercise as a test under equal assignment probabilities, derive confidence sets by test inversion and propose a sensitivity analysis with respect to those probabilities. Against this backdrop, \citet{shaikh2021} develop randomisation tests for staggered adoption. The validity of such tests rests on exchangeability of the treated unit and the donors, which fails when the treated unit is unlike its donors, and \citet{ferman2019} show in the related difference-in-differences setting with few treated groups how heteroskedasticity across units distorts permutation inference and how the test statistic can be rescaled to restore validity. \citet{conley2011} obtain inference with few policy changes by using the untreated units to estimate the distribution of the error of the treated ones, an idea that our donor calibration shares whereas \citet{hahn2017} show that the placebo permutation test controls size only under restrictive conditions on the distribution of the errors.

Time-series permutation is the plausible alternative. For example, \citet{chernozhukov2021} propose a conformal procedure that permutes the residuals of the counterfactual model over time and is exact when those residuals are exchangeable, with moving-block variants for stationary dependence, and \citet{chernozhukov2021t} propose a $t$-test that fits the weights on all but one of $K$ blocks of the pre-treatment period, debiases the estimate with the held-out block and self-normalises over the $K$ fold estimates, which is valid as the numbers of pre-treatment and post-treatment periods grow under stationarity and weak dependence. Their cross-fitting is the closest relative of our time split, and the difference is that we hold out a single later block and condition on the earlier one, which gives an exact finite-sample statement at the price of an explicit bias class. \citet{lei2021} develop conformal inference for individual counterfactuals more generally. \citet{cattaneo2021} construct prediction intervals that separate the uncertainty in the estimated weights, handled by bounds on their deviation from a pseudo-true value, from the out-of-sample uncertainty of the post-treatment error, handled by a quantile model, and they treat the donors as fixed and the treated unit's noise as the source of randomness, which is the conditional view we adopt. \citet{li2020} derives large-sample inference for the average treatment effect with a subsampling procedure, and \citet{bottmer2024} study synthetic control from a design-based perspective in which assignment is random. \citet{viviano2023} obtain model-free inference on treatments over time, and \citet{imai2023} develop matching for time-series cross-sectional data.

What these methods share is that the noise entering the interval is estimated from pre-treatment residuals of weights that were fitted to the same residuals, or that the dependence is handled by exchangeability. Proposition~\ref{prop:dof} shows, for an unrestricted regression analogue, that with as many donors as periods those residuals carry no information about the noise, a heuristic for the constrained synthetic-control fit rather than a theorem about it, and Theorem~\ref{thm:split} shows that separating the periods that fit the weights from the periods that estimate the noise restores a valid finite-sample calibration.

\subsection{Sensitivity Analysis and Bounded Bias}

The sensitivity class of Assumption~\ref{ass:bias} belongs to the literature on inference under bounded departures from an identifying restriction. \citet{manski2018} bound the variation of untreated outcomes across periods to obtain partial identification in a difference-in-differences setting, and \citet{rambachan2023} develop uniformly valid inference for event studies under restrictions on the post-treatment violation of parallel trends, including a smoothness class that bounds the change in the trend and a relative-magnitudes class that bounds it by its pre-treatment size. \citet{roth2022} shows that pre-testing for parallel trends distorts the subsequent inference. Our linear-drift bound is the simplest member of this family, chosen because it is the right scale for a treated unit whose growth may differ from the donors', and the companion paper \citep{companion} studies the resulting sensitivity sets, their breakdown values and the reinterpretation of placebo sets as members of the same family.

The interval we use is the bounded-bias construction of \citet{donoho1994}, \citet{armstrong2018} and \citet{armstrong2020}, who show that adding the worst-case bias to a critical value computed from the folded normal distribution gives the shortest fixed-length interval with honest coverage over a convex class, and of \citet{armstrong2021}, who apply the same calibration to approximate moment conditions. It coincides in form with the interval of \citet{imbens2004} for an identified set of known width, whose uniformity properties \citet{stoye2009} examines. \citet{kolesar2018} and \citet{imbens2019} apply the construction with a scale estimated from the data in regression-discontinuity designs with many observations, and \citet{armstrong2018} discuss the plug-in. Those constructions take the noise scale as known or consistently estimated, which a panel with a handful of validation periods cannot deliver. The paper's contribution to this literature is a finite-sample version with an estimated scale, in which the Student critical value replaces the normal one and the projection that defines the residuals is chosen so that the scale estimate is independent of the estimation error.

\subsection{Sample Splitting}

Splitting the data into a part that selects a model and a part that evaluates it is an old device for restoring valid significance levels after selection \citep{cox1975}, revived for high-dimensional variable selection by \citet{wasserman2009} and studied in the assumption-lean setting by \citet{rinaldo2019}. In the synthetic-control literature, \citet{abadie2015} split the pre-treatment period into a training and a validation block to choose the predictor weights, and cross-validation over time is the standard way to choose penalties \citep{benmichael2021,hollingsworth2022}. These splits serve estimation. The split in this paper serves inference, and its validity rests on the independence of the validation and post-treatment disturbances from the training block rather than on any property of the fitting rule.

\subsection{Addition to the Literature}

Relative to the estimation literature, this paper offers nothing new about how to fit weights, and it covers every fitting rule that uses the training block. But relative to the inference literature, it replaces residual-based or exchangeability-based calibration by an exact conditional calibration under Gaussian time-independent noise that holds with as few as three validation residuals, although four or more are needed for a usable critical value, states the price in degrees of freedom, and gives a conservative finite-sample calibration under bounded serial dependence. Relative to the sensitivity literature, it carries the bounded-bias interval to estimated scales, characterises the anchor that minimises its oracle length, identifies the trend-balancing interval as a classical prediction interval, gives the bias bound and the optimal anchor under a bounded-curvature class with the slope free by a Peano-kernel argument, and proposes the breakdown curvature as the summary of an application. Relative to placebo inference, it shows that the placebo distribution can calibrate a drift bound for the treated unit at a price of at most $1/(J+1)$, with the anchor chosen after the data are seen. Relative to the sample-splitting literature, it shows that a split in time delivers finite-sample validity for data-dependent weights in a panel with one treated unit, which is a setting that the cross-sectional theory does not cover. Each of these contributions is modest on its own. Their combination is a procedure that a practitioner with a dozen pre-treatment periods and a dozen donors can apply, and that states exactly what it assumes.

\section{Theory}\label{sec:theory}
The analysis proceeds from a sampling model and a sensitivity class to an anchored treatment-effect estimator, its finite-sample calibration, and the choice of anchor. Sampling statements condition on the training sample. Proofs are collected in Appendix~\ref{app:proofs}.

\subsection{Setting and Restrictions}\label{sec:setup}

\subsubsection{Potential Outcomes and Assignment}
There are $J+1$ units, indexed by $i\in\{1,\ldots,J+1\}$, and a finite set of observation periods. Unit $1$ is treated in the post-treatment set $\Q$, with $m=\abs{\Q}\ge1$. The pre-treatment periods are partitioned in advance into a training set $\Pone$ and a validation set $\Ptwo$, with $n=\abs{\Ptwo}$. Write $Y_{it}(0)$ for the untreated potential outcome and let
\begin{equation}\label{eq:observed}
 Y_{it}=Y_{it}(0)+D_{it}\tau_t,
 \qquad D_{it}=\mathbf 1\{i=1,\ t\in\Q\}.
\end{equation}
Equation~\eqref{eq:observed} imposes no anticipation and no interference with donor outcomes. The target is the average additive effect
\begin{equation}\label{eq:target}
 \theta=\frac1m\sum_{t\in\Q}\tau_t.
\end{equation}
The treatment effects may vary arbitrarily across post-treatment periods and are fixed in the sampling experiment.

\begin{assumption}[Sampling experiment and assignment]\label{ass:prim}
For each unit and period,
\begin{equation}\label{eq:dgp}
 Y_{it}(0)=\mu_{it}+\varepsilon_{it},
 \qquad \varepsilon_t=(\varepsilon_{1t},\ldots,\varepsilon_{J+1,t})'
 \overset{\mathrm{ind}}{\sim}N(0,\Omega),
\end{equation}
where the mean paths $\{\mu_{it}\}$ are deterministic, $\Omega$ is a fixed positive semidefinite matrix, and the disturbance vectors are independent across periods. The treatment assignment, treatment date, time partition, and deterministic time covariate $x_t$ are fixed. Thus sampling probability refers to the disturbances in~\eqref{eq:dgp}, conditional on the assignment and mean paths.
\end{assumption}

The assignment may be associated with the untreated mean paths. In particular, no random assignment across units is assumed. The Gaussian disturbance law is imposed for the fixed assignment under study. If assignment is instead random, the results require the same conditional disturbance law after conditioning on the realized assignment and training information. Selection on future disturbances does not generally satisfy that condition.

Let $Y_{ct}=(Y_{2t},\ldots,Y_{J+1,t})'$ and let $\F$ be the sigma-field generated by all training outcomes and any independent randomization used by the fitting algorithm. A donor-weight estimator $\widehat w\in\R^J$ must be $\F$-measurable. The usual synthetic-control constraint set is the simplex
\begin{equation}\label{eq:simplex}
 \mathcal W_{\Delta}=\{w\in\R^J:w_j\ge0,\ \one_J'w=1\},
\end{equation}
whose fitted outcome vector lies in the convex hull of the donor outcome vectors. The inference argument also permits other weight sets, provided the selected weights are finite and the assumptions below hold for their realized value. For example, $\widehat w$ may minimize a penalized training loss jointly with an intercept. The penalty, donor selection, and tie-breaking rule must use training information only.

Conditional on $\F$, write $w=\widehat w$ for the realized weights and define
\begin{align}
 u_t(w)&=Y_{1t}-w'Y_{ct}, &
 \delta_t(w)&=\mu_{1t}-w'\mu_{ct},\label{eq:gaps}\\
 \eta_t(w)&=\varepsilon_{1t}-w'\varepsilon_{ct}, &
 \omega^2(w)&=(1,-w')\Omega(1,-w')'.\label{eq:gapvariance}
\end{align}
Here $u_t$ is an observed gap, $\delta_t$ is its untreated mean, and $\eta_t$ is a structural disturbance. Assume $\omega^2(w)>0$ almost surely. Under cross-unit independence, $\omega^2(w)=\omega_1^2+\sum_{j=1}^{J}w_j^2\omega_{j+1}^2$. Unless stated otherwise, all subsequent distributional calculations condition on $\F$ and suppress the argument $w$.

\subsubsection{A Sensitivity Class for Imperfect Fit}
Let $\K\subseteq\Ptwo$ be a prespecified set of potentially contaminated validation periods. More generally, $\K$ may be selected using $\F$. Let $L,M,S\ge0$ be finite sensitivity constants, specified in advance or selected using $\F$ subject to the following almost-sure restrictions.

\begin{assumption}[Counterfactual discrepancy]\label{ass:bias}
For the realized donor weights, there exist $\F$-measurable nuisance quantities $c,d,\ell$ and $\kappa=(\kappa_t:t\in\Ptwo)$ such that, on $\Ptwo\cup\Q$,
\begin{equation}\label{eq:discrepancy}
 \delta_t=c+d x_t+\kappa_t\mathbf1\{t\in\Ptwo\}
                  +\ell\mathbf1\{t\in\Q\},
\end{equation}
where
\begin{equation}\label{eq:sensitivity}
 c\in\R,\qquad |d|\le M,\qquad |\ell|\le L,
 \qquad \kappa_t=0\ (t\notin\K),\qquad |\kappa_t|\le S\ (t\in\K).
\end{equation}
These restrictions hold almost surely under every data-generating process in the maintained class.
\end{assumption}

The unrestricted intercept $c$ permits a persistent level mismatch. The bound $M$ restricts the unmatched untreated trend, $L$ bounds a post-treatment shift in the untreated discrepancy, and $S$ bounds localized validation-period departures. The term $\ell$ is a counterfactual change that is observationally confounded with a constant component of treatment. Its bound is therefore an identifying restriction. The constants are stated in the outcome and time units used in~\eqref{eq:discrepancy}. These restrictions are exact sensitivity restrictions, rather than an asymptotic approximation derived from pre-treatment fit. A small training residual does not establish them. More general mean-path classes can be accommodated by replacing the bias bound below with the support function of the corresponding discrepancy set. The maintained model isolates a tractable class for which both inference and anchor choice admit explicit characterizations. The distinction between fit and counterfactual validity is central to the imperfect-fit problem studied by \citet{ferman2021}.

\subsection{The Anchored Estimator}\label{sec:estimator}
An affine temporal anchor is a vector
\begin{equation}\label{eq:anchor-set}
 \lambda\in\Lambda_n=\{\lambda\in\R^n:\one_n'\lambda=1\}.
\end{equation}
This affine hyperplane has dimension $n-1$ and permits negative weights. Adding $\lambda\ge0$ produces a convex temporal average, but changes the optimization results in Section~\ref{sec:anchor}. Let $\widehat\lambda\in\Lambda_n$ be $\F$-measurable. Conditional on $\F$, write $\lambda=\widehat\lambda$ and set
\begin{equation}\label{eq:estimator}
 \widehat\theta(\lambda)=\bar u_{\Q}-\lambda'u_{\Ptwo},
 \qquad \bar u_{\Q}=\frac1m\sum_{t\in\Q}u_t.
\end{equation}
The intercept discrepancy cancels because $\one_n'\lambda=1$. With $\bar x_{\Q}=m^{-1}\sum_{t\in\Q}x_t$, define
\begin{align}
 h(\lambda)&=\bar x_{\Q}-\lambda'x_{\Ptwo},\label{eq:h}\\
 B(\lambda)&=L+M|h(\lambda)|+S\sum_{t\in\K}|\lambda_t|,\label{eq:B}\\
 a(\lambda)&=\left(m^{-1}+\lambda'\lambda\right)^{1/2},
 \qquad \sigma(\lambda)=\omega a(\lambda).\label{eq:sigma}
\end{align}

\begin{proposition}[Conditional bias decomposition]\label{prop:decomposition}
Under Assumptions~\ref{ass:prim}--\ref{ass:bias},
\begin{equation}\label{eq:decomposition}
 \widehat\theta(\lambda)-\theta
 =\beta(\lambda)+\xi(\lambda),\qquad
 \beta(\lambda)=\ell+d h(\lambda)-\lambda'\kappa,
\end{equation}
where $|\beta(\lambda)|\le B(\lambda)$ and, conditionally on $\F$,
\begin{equation}\label{eq:normal}
 \xi(\lambda)=\bar\eta_{\Q}-\lambda'\eta_{\Ptwo}
 \sim N(0,\sigma^2(\lambda)).
\end{equation}
For fixed $\lambda$, $B(\lambda)$ is the exact supremum of $|\beta(\lambda)|$ over the rectangular nuisance set~\eqref{eq:sensitivity}.
\end{proposition}

Writing $\mu_\lambda=\E[\widehat\theta(\lambda)\mid\F]$, the restriction yields the sensitivity region $[\mu_\lambda-B(\lambda),\mu_\lambda+B(\lambda)]$ for $\theta$ based on this mean contrast. It need not be the sharp identified set obtained from all outcome means. A confidence interval must account for both the discrepancy bound and sampling variation about $\mu_\lambda$.

\subsection{Why Training Fit is Not a Variance Estimate?}\label{sec:insample}
The following auxiliary regression result separates degrees-of-freedom arithmetic from the structural synthetic-control model.

\begin{proposition}[In-sample residual contraction]\label{prop:dof}
Consider $y=Zb_0+v$ on $N$ training observations, where the design $Z$ contains an intercept, has rank $p$, and satisfies
\begin{equation}\label{eq:regressioncondition}
 \E[v\mid Z]=0,\qquad \E[vv'\mid Z]=\omega^2 I_N.
\end{equation}
Then unrestricted least-squares residuals $r=(I_N-P_Z)y$ satisfy
\begin{equation}\label{eq:rss}
 \E[r'r\mid Z]=\omega^2(N-p).
\end{equation}
If $Z$ has $J+1$ columns and rank $\min\{J+1,N\}$, this equals $\omega^2\max\{0,N-J-1\}$. If $p=N$, the unrestricted residual is identically zero. If a constrained fit contains the true slope vector and leaves the intercept unrestricted, its minimized residual sum of squares satisfies
\begin{equation}\label{eq:constrainedrss}
 \omega^2(N-p)\le\E[\operatorname{RSS}_{\mathrm{con}}\mid Z]
 \le\omega^2(N-1).
\end{equation}
The same bounds apply to the infimum when a constrained minimum is not attained.
\end{proposition}

The conditional regression restriction~\eqref{eq:regressioncondition} is essential. In the structural model, a fixed-weight discrepancy contains $\varepsilon_{1t}-w_0'\varepsilon_{ct}$, while the donor design contains $\varepsilon_{ct}$. Under cross-unit independence,
\begin{equation}\label{eq:design-correlation}
 \Cov(\varepsilon_{ct},\varepsilon_{1t}-w_0'\varepsilon_{ct})
 =-\Omega_{cc}w_0,
\end{equation}
which is generally nonzero. Thus~\eqref{eq:rss} cannot be imported into a noisy-donor regression without an additional conditional model. The time-split construction instead conditions on training outcomes and uses independent validation and post-treatment disturbances.

\subsection{Finite-Sample Inference}\label{sec:split}
\subsubsection{Scale Independent Estimate}
Define the validation design and its orthogonal projector by
\begin{equation}\label{eq:X}
 X_\lambda=[\one_n,x_{\Ptwo},\lambda],\qquad
 P_\lambda=X_\lambda(X_\lambda'X_\lambda)^+X_\lambda',\qquad
 \nu_\lambda=n-\rank(X_\lambda),
\end{equation}
where $+$ denotes the Moore--Penrose inverse. Require $\nu_\lambda\ge1$ almost surely. The residual and scale estimators are
\begin{equation}\label{eq:scale}
 e_\lambda=(I_n-P_\lambda)u_{\Ptwo},\qquad
 \widehat\omega_\lambda^2=\frac{e_\lambda'e_\lambda}{\nu_\lambda},\qquad
 \widehat\sigma_\lambda=a(\lambda)\widehat\omega_\lambda.
\end{equation}
Including $\lambda$ in $X_\lambda$ removes the validation-noise direction used by the estimator. Consequently, Gaussian orthogonality delivers independence between the numerator disturbance and the residual scale. If $\lambda\in\col([\one_n,x_{\Ptwo}])$, this extra column costs no additional degree of freedom. With a nonconstant time covariate, at least three validation periods are needed in that case, and at least four when the anchor adds a third independent column.

\begin{lemma}[Orthogonal studentization]\label{lem:student}
Under Assumptions~\ref{ass:prim}--\ref{ass:bias}, conditionally on $\F$,
\begin{equation}\label{eq:noncentral}
 \frac{\nu_\lambda\widehat\omega_\lambda^2}{\omega^2}
 \sim\chi^2_{\nu_\lambda}(\zeta_\lambda),\qquad
 \zeta_\lambda=\frac{\norm{(I_n-P_\lambda)\kappa}^2}{\omega^2},
\end{equation}
and this variable is independent of $Z_\lambda=\xi(\lambda)/\sigma(\lambda)\sim N(0,1)$. If $(I_n-P_\lambda)\kappa=0$, the chi-square distribution is central and
\begin{equation}\label{eq:pivot}
 \frac{\widehat\theta(\lambda)-\theta-\beta(\lambda)}{\widehat\sigma_\lambda}
 \sim t_{\nu_\lambda}.
\end{equation}
For arbitrary allowed $\kappa$, the scale in~\eqref{eq:noncentral} first-order stochastically dominates its central counterpart.
\end{lemma}

\subsubsection{A Feasible Honest Interval and Test}
For $0<\alpha<1/2$, let $t_{\nu,p}$ denote the $p$-quantile of Student's $t_\nu$ distribution. Define
\begin{equation}\label{eq:honest-ci}
 C_{\mathrm{split}}(1-\alpha)=
 \left[\widehat\theta(\lambda)\ \pm\
 \left\{B(\lambda)+t_{\nu_\lambda,1-\alpha/2}\widehat\sigma_\lambda\right\}\right],
\end{equation}
where $[a\pm b]$ denotes $[a-b,a+b]$.

\begin{definition}[Finite-sample honesty]\label{def:honesty}
For a class $\mathfrak P$ of sampling laws and corresponding targets $\theta(P)$, an interval $C$ is honest at level $1-\alpha$ if
\begin{equation}\label{eq:honesty}
 \inf_{P\in\mathfrak P}\Prob_P\{\theta(P)\in C\}\ge1-\alpha.
\end{equation}
The infimum includes all admissible mean paths, disturbance covariance matrices, treatment effects, and nuisance parameters, with the fitting and selection rules fixed.
\end{definition}

\begin{theorem}[Time-split validity]\label{thm:split}
Let $\mathfrak P$ satisfy Assumptions~\ref{ass:prim}--\ref{ass:bias}, training measurability of all selection rules, $\omega^2>0$, and $\nu_\lambda\ge1$. For every $P\in\mathfrak P$,
\begin{equation}\label{eq:conditional-coverage}
 \Prob_P\{\theta(P)\in C_{\mathrm{split}}(1-\alpha)\mid\F\}
 \ge1-\alpha\qquad P\text{-almost surely}.
\end{equation}
Hence~\eqref{eq:honesty} holds without an asymptotic approximation. Equivalently, the test
\begin{equation}\label{eq:test}
 T(\theta_0)=\frac{\pos{|\widehat\theta(\lambda)-\theta_0|-B(\lambda)}}{\widehat\sigma_\lambda},
 \qquad \text{reject if }T(\theta_0)>t_{\nu_\lambda,1-\alpha/2},
\end{equation}
has size at most $\alpha$ uniformly over $\{P\in\mathfrak P:\theta(P)=\theta_0\}$. Here $(a)_+=\max\{a,0\}$.
\end{theorem}

The theorem permits many donor units and arbitrary training algorithms because the held-out disturbances are independent of the selected weights. Its finite-sample calibration still relies on Gaussianity, time independence, and a common gap variance across validation and post-treatment periods. A time split alone does not establish these distributional properties.

\begin{remark}[Calibration and selection]\label{rem:calibration}
Sensitivity constants may be fixed externally or computed from training information if Assumption~\ref{ass:bias} holds almost surely. If the bounds hold only on a training-measurable event of probability at least $1-\gamma$ uniformly, using nominal error probability $\alpha_0$ gives coverage at least $1-\alpha_0-\gamma$. Thus $\alpha_0=\alpha-\gamma$, with $0\le\gamma<\alpha$, suffices for unconditional coverage $1-\alpha$. Selection of the anchor or sensitivity specification using validation or post-treatment disturbances is not covered by the conditional theorem. In particular, maximizing a donor-placebo discrepancy does not itself certify a bound for the treated unit, and Theorem~\ref{thm:exch} gives the additional condition under which it does.
\end{remark}

\begin{corollary}[Deterministic donors]\label{cor:don}
If every donor disturbance has zero variance, functions of donor outcomes alone, including post-treatment donor outcomes, are deterministic in the maintained experiment. They may therefore enter the sensitivity specification without compromising the conditioning argument, provided any placebo pool excludes the treated unit and Assumption~\ref{ass:bias} remains valid. The latter restriction is a substantive requirement, independent of measurability.
\end{corollary}

\subsubsection{Known-Scale and Oracle Calibrations}
For $b\ge0$, define $f_\alpha(b)=\cv(b)$ as the unique positive solution of
\begin{equation}\label{eq:folded}
 \Phi(f_\alpha(b)-b)-\Phi(-f_\alpha(b)-b)=1-\alpha,
\end{equation}
and define the bias-aware Gaussian radius
\begin{equation}\label{eq:chi}
 \chi_\alpha(B,\sigma)=\sigma f_\alpha(B/\sigma).
\end{equation}
Here $\Phi$ and $\varphi$ denote the standard normal distribution and density functions. This calibration is the scalar bounded-bias construction used in honest confidence intervals and approximate-moment inference, as developed by \citet{armstrong2020,armstrong2021}.

\begin{proposition}[Known-scale calibration]\label{prop:oracle}
If $\sigma(\lambda)$ is known conditionally on $\F$, then
\begin{equation}\label{eq:gaussian-ci}
 C_G(1-\alpha)=\left[\widehat\theta(\lambda)\ \pm\
 \chi_\alpha(B(\lambda),\sigma(\lambda))\right]
\end{equation}
has conditional coverage at least $1-\alpha$. Coverage equals $1-\alpha$ at $\beta(\lambda)=\pm B(\lambda)$. In the scalar experiment $\widehat\theta=\theta+\beta+\sigma Z$, $|\beta|\le B$, this is the smallest deterministic symmetric half-length around the specified estimator that attains uniform coverage $1-\alpha$.
\end{proposition}

For comparison, when $\zeta_\lambda=0$, let $W\sim\sqrt{\chi^2_\nu/\nu}$ be independent of $Z\sim N(0,1)$ and define $k_\alpha(r,\nu)$ by
\begin{equation}\label{eq:k-calibration}
 \E\!\left[\Phi(k_\alpha(r,\nu)W)-\Phi(-2r-k_\alpha(r,\nu)W)\right]=1-\alpha,
 \qquad r\ge0.
\end{equation}
The interval $[\widehat\theta\pm\{B+k_\alpha(B/\sigma,\nu)\widehat\sigma\}]$ has exact least-favourable coverage when the ratio $B/\sigma$ is known. The critical value is nonincreasing in $r$, with
\begin{equation}\label{eq:k-limits}
 k_\alpha(0,\nu)=t_{\nu,1-\alpha/2},\qquad
 \lim_{r\to\infty}k_\alpha(r,\nu)=t_{\nu,1-\alpha}.
\end{equation}
This is an oracle calibration when the scale is unknown and $B>0$. Substitution of $B/\widehat\sigma$ is not justified by the preceding independence argument. Equation~\eqref{eq:honest-ci} remains the feasible finite-sample procedure, and it is least favourable in the following sense. The coverage of $[\widehat\theta\pm(B+k\widehat\sigma)]$ at the boundary of the bias envelope is $\Prob(-2r-kW\le Z\le kW)$ with $r=B/\sigma$, which is increasing in $r$, and since $\sigma$ is unrestricted the infimum over the class is the limit $r\to0$, where it equals $\Prob(|Z|\le kW)$ and forces $k=t_{\nu,1-\alpha/2}$. A two-stage sharpening that replaces $r$ by a lower confidence bound $B/\sigma_U$, with $\sigma_U$ an upper confidence bound on $\sigma$ at level $1-\alpha_1$, and uses $k_{\alpha-\alpha_1}(B/\sigma_U,\nu)$, is valid by a union bound but gains little at the degrees of freedom this paper is about. With $\nu=4$ and $\alpha_1=0.005$, $\sigma_U=4.4\widehat\sigma$, so $r$ is replaced by a quarter of its plug-in value, and the critical value at level $0.045$ exceeds $t_{4,0.975}$ whenever $B/\widehat\sigma$ is below about $0.26$ and is at most a fifth smaller than it otherwise.

\subsubsection{Power in the Gaussian Experiment}
\begin{proposition}[A worst-case power bound]\label{prop:power}
For the known-scale test that rejects $\theta=\theta_0$ when $\theta_0\notin C_G$, write $\Delta=\theta-\theta_0$, $r=B/\sigma$, $c_\alpha=f_\alpha(r)$, and $d_\Delta=(|\Delta|-B)_+/\sigma$. Its rejection probability satisfies
\begin{equation}\label{eq:power}
 \Prob\{\theta_0\notin C_G\mid\F\}
 \ge1-\Phi(c_\alpha-d_\Delta)+\Phi(-c_\alpha-d_\Delta).
\end{equation}
The bound is attained by minimizing over the scalar nuisance interval $\beta\in[-B,B]$. Along any sequence of such conditional Gaussian experiments, power converges to one if $(|\Delta|-2B)/\sigma\to\infty$.
\end{proposition}
The separation condition exhibits the effect of counterfactual uncertainty on detection. A nonvanishing bias envelope can prevent informative testing even when sampling variance becomes small. The proposition concerns the known-scale experiment and does not assert the same power for an estimated scale inflated by validation contamination.

\subsection{The Choice of Anchor}\label{sec:anchor}
\subsubsection{Conditional Fixed-Length Optimization}
This section studies a known-scale Gaussian design problem conditional on the training sample. To allow a more general design benchmark, suppose the validation gap disturbance has covariance $\omega^2R$, where $R\in\R^{n\times n}$ is known and positive definite, and is independent of the post-treatment average disturbance, whose variance is $\omega^2/m$. Set
\begin{equation}\label{eq:sigmaR}
 \sigma_R(\lambda)=\omega\left(m^{-1}+\lambda'R\lambda\right)^{1/2}.
\end{equation}
Assumption~\ref{ass:prim} is the special case $R=I_n$. For $R\ne I_n$, the scale construction~\eqref{eq:scale} would require the whitening modification described in Remark~\ref{rem:R} below.

Initially set $S=0$ and consider
\begin{equation}\label{eq:optimization}
 \lambda^*\in\argmin_{\lambda\in\Lambda_n}
 F(\lambda),\qquad
 F(\lambda)=\chi_\alpha\!\left(L+M|\bar x_{\Q}-\lambda'x_{\Ptwo}|,\sigma_R(\lambda)\right).
\end{equation}
The criterion minimizes the fixed confidence radius for the stated affine-estimator class and bias envelope. This is a conditional length-optimality result. A minimax-regret claim over unrestricted procedures would require a separate decision problem and lower bound.

\begin{lemma}[Geometry of the critical radius]\label{lem:cv}
For $0<\alpha<1/2$, $f_\alpha$ is increasing and strictly convex on $[0,\infty)$, with
\begin{align}
 f_\alpha'(b)&=\tanh(b f_\alpha(b)),\label{eq:fprime}\\
 f_\alpha''(b)&=\{1-\tanh^2(b f_\alpha(b))\}
                       \{f_\alpha(b)+b f_\alpha'(b)\}>0.
\end{align}
Moreover, $f_\alpha(0)=z_{1-\alpha/2}$ and $f_\alpha(b)-b\downarrow z_{1-\alpha}$. The map $\chi_\alpha$ is jointly convex, nondecreasing in $B$, and strictly increasing in $\sigma>0$. With $b=B/\sigma$,
\begin{equation}\label{eq:chi-derivatives}
 \partial_B\chi_\alpha=f_\alpha'(b),\qquad
 \partial_\sigma\chi_\alpha=g_\alpha(b)
 :=f_\alpha(b)-b f_\alpha'(b),
\end{equation}
where $z_{1-\alpha}<g_\alpha(b)\le z_{1-\alpha/2}$. Finally,
\begin{equation}\label{eq:chi-bounds}
 B+z_{1-\alpha}\sigma\le\chi_\alpha(B,\sigma)
 \le B+z_{1-\alpha/2}\sigma.
\end{equation}
\end{lemma}

Assume $A=[\one_n,x_{\Ptwo}]$ has rank two, and define
\begin{align}
 G&=(A'R^{-1}A)^{-1}, & v(a)&=(1,a)',\label{eq:G}\\
 q(a)&=v(a)'Gv(a), &
 a_0&=\frac{\one_n'R^{-1}x_{\Ptwo}}{\one_n'R^{-1}\one_n},\label{eq:q}\\
 h_0&=\bar x_{\Q}-a_0, &
 s(h)&=\omega\{m^{-1}+q(\bar x_{\Q}-h)\}^{1/2}.\label{eq:sh}
\end{align}

\begin{theorem}[Optimal affine anchor]\label{thm:affine}
Suppose $R\succ0$, $\omega>0$, $\rank(A)=2$, and $0<\alpha<1/2$.
\begin{enumerate}[label=(\roman*),leftmargin=2em]
\item The objective in~\eqref{eq:optimization} is continuous, strictly convex, and coercive on $\Lambda_n$. It has a unique minimizer.
\item For fixed $h$, the optimal anchor and reduced objective are
\begin{equation}\label{eq:lambda-h}
 \lambda(h)=R^{-1}AGv(\bar x_{\Q}-h),\qquad
 \phi(h)=\chi_\alpha(L+M|h|,s(h)).
\end{equation}
Thus $\lambda^*=\lambda(h^*)$, where $h^*$ uniquely minimizes $\phi$ on $\R$. The minimizer lies between $0$ and $h_0$, regardless of the sign of $h_0$.
\item If $M=0$, the solution is the minimum-variance anchor
\begin{equation}\label{eq:mv-anchor}
 \lambda^*=\frac{R^{-1}\one_n}{\one_n'R^{-1}\one_n},\qquad h^*=h_0.
\end{equation}
For $R=I_n$, this is the uniform pre-treatment average. If $h_0=0$, the same anchor is optimal for every $L,M\ge0$.
\item For the remaining statements, orient $x$ so that $h_0>0$. If $L=0$ and $M>0$, then $0<h^*<h_0$. The optimal mismatch is nonincreasing in $M$ and converges to zero as $M\to\infty$.
\item If $L>0$, put $b_0=L/s(0)$. Exact trend balance, $h^*=0$, occurs if and only if
\begin{equation}\label{eq:M-threshold}
 M\ge M_{\mathrm{bal}}
 :=\frac{-g_\alpha(b_0)s'(0)}{f_\alpha'(b_0)},\qquad
 s'(0)=-\frac{\omega^2(G_{12}+G_{22}\bar x_{\Q})}{s(0)}<0.
\end{equation}
When $R=I_n$, $G_{12}+G_{22}\bar x_{\Q}=(\bar x_{\Q}-\bar x_{\Ptwo})/\sum_{t\in\Ptwo}(x_t-\bar x_{\Ptwo})^2$.
\end{enumerate}
\end{theorem}

The theorem describes a variance--bias trade-off within the affine anchor class. With no post-treatment level allowance, a finite trend bound leaves some trend mismatch. A positive level allowance can instead make exact trend balance optimal at a finite threshold. These statements concern oracle length at the true scale. A training estimate of scale can be used to select an implementable anchor without invalidating Theorem~\ref{thm:split}, but it does not inherit oracle length optimality automatically.

\begin{remark}[Known covariance shape]\label{rem:R}
Under the conditional covariance model used in this section, take the symmetric square root of $R$ and define
\begin{equation}\label{eq:whiten}
 \widetilde u=R^{-1/2}u_{\Ptwo},\qquad
 \widetilde X=[R^{-1/2}\one_n,R^{-1/2}x_{\Ptwo},R^{1/2}\lambda].
\end{equation}
Replacing the residual in~\eqref{eq:scale} with $(I-P_{\widetilde X})\widetilde u$, the degrees of freedom with $n-\rank(\widetilde X)$, and $a(\lambda)$ with $(m^{-1}+\lambda'R\lambda)^{1/2}$ recovers the same proof. This extension requires $R$ to be known and the conditional Gaussian block-independence assumptions to hold. Estimating a serial covariance matrix from the validation outcomes is outside this exact argument.
\end{remark}

\begin{figure}[t]
\centering
\includegraphics[width=\textwidth]{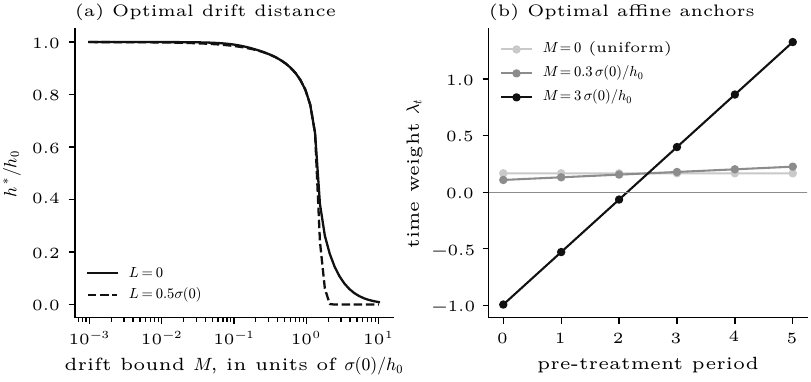}
\caption{The optimal affine anchor}\label{fig:anchor}
\fignote{Six pre-treatment periods, ten post-treatment periods, independent noise. Panel (a) plots the optimal drift distance $h^{*}$ relative to the minimum-variance distance $h_0$ against the drift bound $M$ measured in units of $s(0)/h_0$, written $\sigma(0)$ in the panels, for $L=0$ and for a level bound of half the trend-balancing anchor's noise. Panel (b) plots the time weights of the optimal anchor for three values of $M$.}
\end{figure}

Figure~\ref{fig:anchor} illustrates Theorem~\ref{thm:affine}. Under any positive drift bound the uniform anchor is dominated, however small the bound, because moving the anchor toward the post-treatment period reduces the drift distance at first order while raising the noise only at second order near the minimum-variance anchor, and the gain is small when $M$ is small relative to the noise. Trend extrapolation is never exactly optimal when the level shift is excluded, because the critical value is flat in the bias at zero bias, so a small drift exposure is free at first order while the extrapolation's noise is not. In the split construction the scale that the optimiser needs must be training-measurable for Theorem~\ref{thm:split} to apply to the selected anchor. The treated unit's training residuals are the obvious candidate and, by Proposition~\ref{prop:dof}, a bad one. The implementation pools the donors' leave-one-period-out prediction errors on the training block, net of a line, which uses no validation outcome and is informative when the donors are as noisy as the treated unit. Theorem~\ref{thm:exch} below gives a second route, in which the anchor may be chosen with the validation scale itself at the price of a simultaneous critical value.

Theorem~\ref{thm:affine} concerns a fixed-length interval over an affine class in the sense of \citet{donoho1994} and \citet{armstrong2018}, with the post-treatment weights frozen at $m^{-1}\one_m$, the optimisation confined to the anchor and the scale known. It is a statement of oracle optimality within the anchored class and not a minimax or efficiency claim, since the efficiency bounds of those papers refer to all affine estimators with known variance. The interval of \citet{rambachan2023} for event studies is the same construction with the post-treatment weights free, and freeing them here would shorten the interval when the post-treatment periods are unequally informative, at the cost of the simple average as the target. We keep the average because it is the estimand of the applied literature and because the one-dimensional reduction of part (ii) is what makes the anchor interpretable.

\begin{remark}[The feasible criterion]\label{rem:feasible}
Theorem~\ref{thm:affine} and Propositions~\ref{prop:exclusion} and~\ref{prop:soft} minimise the oracle radius $\chi_\alpha(B,\sigma_R)$, whose flatness in the bias at zero, $f_\alpha'(0)=0$, drives the results that trend balance is never exactly optimal when $L=0$ and that contaminated periods keep positive weight at every finite $S$. The feasible interval \eqref{eq:honest-ci} adds the bias to a Student radius, and for the additive criterion $\psi(h)=L+M|h|+t\,s(h)$ these two conclusions reverse. With $R=I_n$ and $h_0>0$, $\psi$ is convex with right derivative $M+t\,s'(0)$ at zero, so trend balance is optimal if and only if $M\ge-t\,s'(0)$ whatever $L$, and in the design of Proposition~\ref{prop:soft} the contaminated periods are excluded exactly once $S\ge-t\,s_c'(0)$. The oracle results describe the known-scale benchmark, and an implementation that targets the feasible length should use $\psi$. The simulations select the anchor for the linear class by a third criterion, the quantile of a folded noncentral $t$ at the training-block scale. Any training-measurable criterion leaves Theorem~\ref{thm:split} intact, and the choice affects only length.
\end{remark}

\begin{remark}[The trend-balancing interval is a prediction interval]\label{rem:pi}
With $R=I_n$, the trend-balancing anchor $\lambda_{\mathrm{tr}}=\lambda(0)=A(A'A)^{-1}(1,\bar x_{\Q})'$ gives $\lambda_{\mathrm{tr}}'u_{\Ptwo}=\widehat a+\widehat b\,\bar x_{\Q}$, where $(\widehat a,\widehat b)$ are the least-squares coefficients of the validation-block gap on $(1,x_t)$, and $\sigma^2(\lambda_{\mathrm{tr}})=\omega^2\{m^{-1}+n^{-1}+(\bar x_{\Q}-\bar x_{\Ptwo})^2/\sum_{\Ptwo}(x_t-\bar x_{\Ptwo})^2\}$ with $\nu=n-2$. The interval of Theorem~\ref{thm:split} with $L=S=0$ is therefore the classical prediction interval for the mean of $m$ future observations from a line fitted to the validation-block gap. What the time split adds to that textbook object is the condition under which it is valid when the gap is built from weights estimated on the same panel, namely that the weights use only the training block.
\end{remark}

\begin{remark}[Unequal variances across blocks]\label{rem:hetero}
Assumption~\ref{ass:prim} gives the validation and post-treatment disturbances the same variance. If the post-treatment variance is instead $\rho^2\omega^2$ with $\rho$ unknown but $\rho\le\bar\rho$ for a known $\bar\rho$, then $\Var(\xi(\lambda)\mid\F)=\omega^2(\rho^2m^{-1}+\lambda'\lambda)\le\omega^2\bar a(\lambda)^2$ with $\bar a(\lambda)=(\bar\rho^2m^{-1}+\lambda'\lambda)^{1/2}$, the validation residuals are unaffected, and $\xi(\lambda)/\{\widehat\omega_\lambda\bar a(\lambda)\}$ is a Gaussian variable of variance at most one divided by an independent $(\chi^2_{\nu_\lambda}/\nu_\lambda)^{1/2}$, so the interval with $\widehat\sigma_\lambda$ replaced by $\widehat\omega_\lambda\bar a(\lambda)$ has coverage at least $1-\alpha$, exactly when the validation residuals are central and by the dominance step of the proof of Theorem~\ref{thm:split} otherwise. The price is the ratio $\bar a(\lambda)/a(\lambda)$, which is small when $\lambda'\lambda$ dominates $m^{-1}$, as it does for the anchors of Section~\ref{sec:anchor} with many post-treatment periods, and which is $\bar\rho$ at most when $\bar\rho\ge1$. A post-treatment variance that is smaller than the validation variance is covered without any change. The remark covers a known bound on the ratio of two block variances under the Gaussian model and does not adapt the oracle calibration or the anchor optimiser, which would use $\bar a$ in place of $a$ if the worst case over $\rho$ were the target.
\end{remark}

\subsubsection{Potentially Contaminated Anchor Periods}
Restore $S\ge0$, let $b_0(\lambda)=L+M|h(\lambda)|$, and consider
\begin{equation}\label{eq:contaminated-objective}
 F_S(\lambda)=\chi_\alpha\!\left(b_0(\lambda)+S\norm{\lambda_{\K}}_1,\sigma_R(\lambda)\right).
\end{equation}
The strict-convexity and existence conclusions remain valid. Suppose $\K$ is a proper subset of the validation periods and let $\lambda^\circ$ minimize $F_0$ over $\Lambda_n\cap\{\lambda_{\K}=0\}$.

\begin{proposition}[A sufficient condition for exact exclusion]\label{prop:exclusion}
If $b_0(\lambda^\circ)>0$, there exists a finite $S_0$ such that the unique minimizer of $F_S$ equals $\lambda^\circ$ for every $S\ge S_0$. More precisely, put $b^\circ=b_0(\lambda^\circ)/\sigma_R(\lambda^\circ)$ and choose $v\in\partial b_0(\lambda^\circ)$ and a scalar $a$ satisfying the restricted optimality conditions
\begin{equation}\label{eq:restricted-KKT}
 g_i+a=0\quad(i\notin\K),\qquad
 g=f_\alpha'(b^\circ)v+g_\alpha(b^\circ)\nabla\sigma_R(\lambda^\circ).
\end{equation}
For nonempty $\K$, one may take
\begin{equation}\label{eq:S-threshold}
 S_0=\frac{\max_{i\in\K}|g_i+a|}{f_\alpha'(b^\circ)}.
\end{equation}
If $\K$ is empty, exclusion is vacuous.
\end{proposition}

The condition concerns the residual bias bound at the restricted optimum. The weaker restriction $L+M>0$ is insufficient in general. For example, take $R=I_3$, $x_{\Ptwo}=(-1,0,1)'$, $\bar x_{\Q}=0$, $\K=\{2\}$, $L=0$, and $M>0$. The optimal anchor is symmetric in the two uncontaminated periods, so it has exact trend balance for every total contaminated weight. The problem therefore reduces to the pure-contamination problem below, whose solution assigns positive contaminated weight for every finite $S>0$.

\begin{proposition}[Pure contamination under independent validation noise]\label{prop:soft}
Suppose $L=M=0$, $R=I_n$, $k=|\K|\in\{1,\ldots,n-1\}$, and $n_0=n-k$. For every finite $S>0$, the optimal anchor assigns a strictly positive total weight $q_S$ to the contaminated periods. Its weights are $(1-q_S)/n_0$ on each uncontaminated period and $q_S/k$ on each contaminated period, where
\begin{equation}\label{eq:soft}
 0<q_S<\frac{k}{n},\qquad
 q_S=\frac{\omega^2}{n_0S^2}+o(S^{-2})\quad\text{as }S\to\infty.
\end{equation}
\end{proposition}
The rate in~\eqref{eq:soft} is a result for this symmetric independent-noise design. Under a general positive-definite $R$, the unrestricted minimum-variance anchor can already have zero contaminated coordinates, so universal strictly positive contamination weights cannot be asserted.

\subsection{Curvature and the Breakdown Curvature}\label{sec:curvature}

The class of Assumption~\ref{ass:bias} is linear in time between the validation block and the post-treatment period, so an anchor with $h(\lambda)=0$ carries no drift bias whatever $M$, and the trend-balancing anchor extrapolates the validation-block slope exactly. Over a post-treatment window of a decade that is a strong restriction, and Section~\ref{sec:abadie} shows that much of the precision of the classic applications rests on it. The class is therefore enlarged by a bounded second derivative, in the spirit of the smoothness class of \citet{rambachan2023}, and the slope is left unrestricted.

\begin{assumption}[Curvature]\label{ass:curv}
Assumption~\ref{ass:bias} holds with $M=\infty$ and with the drift term $dx_t$ replaced by $dx_t+q(x_t)$, where $q$ has an absolutely continuous derivative on the convex hull of the observation times, $q(x_0)=q'(x_0)=0$ at the last validation time $x_0=\max_{t\in\Ptwo}x_t$, and $|q''(x)|\le C$ almost everywhere for a known $C\ge0$.
\end{assumption}

The normalisation at $x_0$ makes $d$ the slope of the discrepancy at the end of the validation block and $C$ the rate at which that slope may change, in outcome units per squared period. With $d$ unrestricted the bias of $\widehat\theta(\lambda)$ is unbounded unless $h(\lambda)=0$, so the admissible anchors are the trend-balanced ones, $\Lambda^0_n=\{\lambda\in\Lambda_n:\lambda'x_{\Ptwo}=\bar x_{\Q}\}$. Define, for $s$ in the hull,
\begin{align}
K(x,s)&=(x-s)_+\,\one\{s\ge x_0\}+(s-x)_+\,\one\{s<x_0\},\label{eq:peano}\\
\kappa_C(\lambda)&=\int\big|k_\lambda(s)\big|\,ds,\qquad
k_\lambda(s)=m^{-1}\sum_{t\in\Q}K(x_t,s)-\sum_{t\in\Ptwo}\lambda_tK(x_t,s).\nonumber
\end{align}

\begin{proposition}[Bias under curvature]\label{prop:curv}
Under Assumption~\ref{ass:curv}, for every $\lambda\in\Lambda^0_n$,
\[
\big|\E[\widehat\theta(\lambda)\mid\F]-\theta\big|\le B_C(\lambda)=L+C\kappa_C(\lambda)+S\sum_{t\in\K}|\lambda_t| ,
\]
and the bound is attained by $q''(s)=C\operatorname{sign}\{k_\lambda(s)\}$ together with the extremal values of the other components. The coverage statements of Theorem~\ref{thm:split} and of Proposition~\ref{prop:dependence} hold with $B(\lambda)$ replaced by $B_C(\lambda)$, since the validation residuals now contain the fixed shift $(I_n-P_\lambda)q(x_{\Ptwo})$ and the scale is inflated rather than exact. The numerical calibration \eqref{eq:k-num} is not covered, for the reason given after it in Section~\ref{sec:dependence}. The function $\kappa_C$ is convex in $\lambda$, its discretisation on a grid containing the observation times bounds it from above, and for fixed degrees of freedom the discretised Student length $B_C(\lambda)+t_{\nu,1-\alpha/2}\,\omega a(\lambda)$ is minimised over $\Lambda^0_n$ by a second-order cone programme. If the slope is instead bounded, $|d|\le M$, the bound acquires the term $M|h(\lambda)|$ and the programme runs over $\Lambda_n$.
\end{proposition}

For $s>x_0$ only the post-treatment terms of $k_\lambda$ survive, so $\kappa_C(\lambda)=(2m)^{-1}\sum_{\Q}(x_t-x_0)^2+\int_{s<x_0}|\sum_{\Ptwo}\lambda_t(s-x_t)_+|\,ds$. The first term is the curvature exposure of the extrapolation itself, which no anchor removes, and the second the exposure of the anchor, which is smallest when the anchor concentrates on the last validation periods. The curvature-optimal anchor therefore trades the variance of a concentrated anchor against the exposure of a spread one, and it is training-measurable whenever the reference scale $\omega$ in the programme is. The implementation uses the scale of the donors' leave-one-period-out prediction errors on the training block, which uses no validation outcome of any unit, so Theorem~\ref{thm:split} applies to the selected anchor exactly. Because the degrees of freedom drop from $n-2$ to $n-3$ when the anchor leaves the span of the intercept and the trend, the implementation compares the cone-programme solution, charged $t_{n-3}$, with the trend-balancing anchor at $t_{n-2}$, both at the training-block scale, and keeps the shorter.

\begin{definition}[Breakdown curvature]\label{def:breakdown}
For a rule that selects a training-measurable anchor $\lambda_C\in\Lambda^0_n$ for each $C\ge0$, the breakdown curvature is $C^*=\inf\{C\ge0:0\in C_{\mathrm{split}}(1-\alpha;C)\}$, where $C_{\mathrm{split}}(1-\alpha;C)$ is the interval of Theorem~\ref{thm:split} with bias bound $B_C(\lambda_C)$. For the trend-balancing anchor $\lambda_{\mathrm{tr}}$ with $L=S=0$,
\begin{equation}\label{eq:Cstar}
C^*_{\mathrm{tr}}=\frac{\big(|\widehat\theta(\lambda_{\mathrm{tr}})|-t_{\nu,1-\alpha/2}\,\widehat\sigma_{\lambda_{\mathrm{tr}}}\big)_+}{\kappa_C(\lambda_{\mathrm{tr}})}.
\end{equation}
\end{definition}

Each interval in the definition is valid at its own $C$, so zero is excluded for every $C<C^*$ and $C^*$ is the smallest curvature bound under which the data are compatible with a zero effect at level $1-\alpha$. Because the selected anchor changes with $C$, the set of bounds at which zero is excluded need not be an interval, and the applications check that it is on a grid. It plays the role of the breakdown values of \citet{rambachan2023} and \citet{manski2018}, in the units of the outcome per squared period. The class itself is related to the smoothness class of \citet{rambachan2023}. At integer times, $|q''|\le C$ implies that every second difference of $q$ is at most $C$ in absolute value, so Assumption~\ref{ass:curv} restricted to the observation times is contained in their class with bound $C$, and the Peano kernel handles unequally spaced and excluded periods. The difference from their interval is that the post-treatment weights are fixed at $m^{-1}\one_m$ and the scale is estimated from validation residuals with a Student calibration. A reference for its size is the curvature the donors display. For donor $j$, let $C_j$ be twice the quadratic coefficient of a quadratic fitted to its placebo gap over the validation and post-treatment periods, and let $C_{\mathrm{don}}=\max_jC_j$. The count of placebos with $C_j>C^*$ is a sensitivity statement and not a guarantee, because the $C_j$ are estimated from noisy gaps and the treated unit's own curvature after treatment is not observed. The simulations of Section~\ref{sec:curvsim} show that $C_{\mathrm{don}}$ is a valid bound when the treated unit bends no more than its donors and no restriction at all otherwise.

\subsection{Donor-Based Drift Bound Calibration}\label{sec:exch}

The drift bound $M$ is the restriction on which the uniform and optimal anchors rest, and a bound computed from the donors' placebo errors is the natural way to set it. Such a bound uses the donors' validation and post-treatment outcomes, so it is not training-measurable and Remark~\ref{rem:calibration} does not cover it. This subsection gives a condition across units under which a placebo-calibrated bound is valid, and states its price.

For each donor $j\in\{2,\dots,J+1\}$, fit weights on the training block to the other donors, with the treated unit excluded, and let $\Pi_j$ be the uniform-anchor split estimate for donor $j$, which estimates a zero effect. For the treated unit let $\Pi_1=\widehat\theta(\lambda_U)-\theta$, the unobserved error of its uniform-anchor estimate, with $\lambda_U=n^{-1}\one_n$ and $h_U=h(\lambda_U)\ne0$. Write $W=\max_{j\ge2}|\Pi_j|$.

The statement needs one probability space for the units and for the noise. Let the mean paths and the assignment be random, drawn from some joint law, and let Assumption~\ref{ass:prim} hold conditionally on them and on $\F$.

\begin{assumption}[Placebo exchangeability]\label{ass:exch}
Under the joint law of the paths and the disturbances, $\Prob\{|\Pi_1|>W\}\le1/(J+1)$.
\end{assumption}

The assumption holds when $(|\Pi_1|,\dots,|\Pi_{J+1}|)$ is exchangeable, which is the content of the placebo test of \citet{abadie2010} as formalised by \citet{firpo2018}, here applied to the untreated error of a split estimator. It is a primitive restriction on the constructed statistics and not a consequence of exchangeability of the units, because $\Pi_1$ is computed from a pool of $J$ donors and each $\Pi_j$ from a pool of $J-1$ with the treated unit excluded. It is a restriction across units, whereas Assumptions~\ref{ass:prim} and~\ref{ass:bias} restrict the treated unit alone, and the guarantee it delivers is unconditional rather than conditional on $\F$. It fails when the treated unit drifts more than every donor.

Consider the anchors $\lambda(h)$, $h\in[0,h_U]$, of part (ii) of Theorem~\ref{thm:affine} with $R=I_n$, which run from trend balance at $h=0$ to the uniform anchor at $h=h_U$. Each lies in $\col([\one_n,x_{\Ptwo}])$, so all of them share the residual projection onto that span, one scale estimate $\widehat\omega$ and $\nu=n-2$ degrees of freedom. Let $k^{\mathrm{sim}}_\alpha$ be the $1-\alpha$ quantile of $\sup_{h\in[0,h_U]}|\xi(\lambda(h))|/\widehat\sigma_{\lambda(h)}$ under Assumption~\ref{ass:prim} with $\kappa=0$. The law of this supremum depends only on $x_{\Ptwo}$, $\bar x_{\Q}$ and $m$, and Monte Carlo computes the quantile to any precision.

\begin{theorem}[Placebo-calibrated drift bound]\label{thm:exch}
Let Assumptions~\ref{ass:prim}, \ref{ass:bias} with $S=0$ and~\ref{ass:exch} hold, with the treated unit's weights training-measurable. Put
\begin{equation}\label{eq:Mhat}
\widehat M=\frac{W+L+k^{\mathrm{sim}}_\alpha\widehat\sigma_{\lambda_U}}{|h_U|},\qquad
C_{\mathrm{exch}}(h)=\Big[\widehat\theta(\lambda(h))\ \pm\ \big\{L+\widehat M|h|+k^{\mathrm{sim}}_\alpha\widehat\sigma_{\lambda(h)}\big\}\Big].
\end{equation}
Then $\Prob\{\theta\in C_{\mathrm{exch}}(\widehat h)\}\ge1-\alpha-1/(J+1)$ under the joint law, for every random $\widehat h$ between $0$ and $h_U$, including the $\widehat h$ that minimises the length of $C_{\mathrm{exch}}$ after the validation outcomes have been seen.
\end{theorem}

The theorem converts the placebo distribution into a drift bound for the treated unit, adds the treated unit's own noise to it, which the raw placebo bound $W/|h_U|$ omits, and lets the anchor be chosen from all the data because the critical value is simultaneous over the family. Its price is at most $1/(J+1)$, which is $0.067$ with fourteen donors and $0.026$ with thirty-eight, so a guarantee of $0.95$ is out of reach with fewer than twenty donors, together with the simultaneous critical value, $3.48$ against $t_{4,0.975}=2.78$ in the simulation design. The trend-balancing member $h=0$ does not use $\widehat M$, and the trend-balancing interval of Theorem~\ref{thm:split} needs neither a drift bound nor exchangeability, so the theorem shortens the interval only when the placebo errors are small relative to the noise, which they are not in the applications of Section~\ref{sec:abadie}. Its main use is to say when the common practice of bounding the drift by the largest placebo error is valid, and the raw bound $W/|h_U|$ is not, since it omits the treated unit's own noise. A placebo-calibrated curvature bound would need the analogous statement for the treated unit's curvature after treatment, which the validation block does not identify, so $C_{\mathrm{don}}$ remains a reference for the breakdown curvature rather than a guarantee.

\section{Bounded Serial Dependence}\label{sec:dependence}

Time independence is the assumption that a short panel cannot check, and Remark~\ref{rem:R} handles only the case in which the dependence is known. This section gives a finite-sample calibration that is valid under any Gaussian dependence whose strength is bounded, at a price in length that the section states. Keep the maintained setting of Section~\ref{sec:setup} with its deterministic mean paths, $\omega^2>0$, training-measurable $w$, $\lambda$, $\K$ and $\kappa$, and $\nu_\lambda\ge1$, but replace the independence part of Assumption~\ref{ass:prim} as follows. Conditionally on $\F$, the gap disturbances $\eta$ on $\Ptwo\cup\Q$ are Gaussian with mean zero and covariance $\omega^2\bar R$, where $\bar R$ is a correlation matrix whose eigenvalues lie in a known interval $[\underline\rho,\overline\rho]$ with $0<\underline\rho\le1\le\overline\rho<\infty$. The conditional statement is the substantive restriction. It holds when the dependence is confined to $\Ptwo\cup\Q$, or when the disturbances of those periods are independent of the training disturbances, and it fails for a process whose memory reaches back into the training block, because the conditional mean of such a process given $\F$ is a decaying function of the last training disturbance that the discrepancy class of Assumption~\ref{ass:bias} does not contain. An excluded block between $\Pone$ and $\Ptwo$ that is long relative to the memory of the process makes the restriction a good approximation rather than an exact one. For a gap process that is a stationary first-order autoregression with coefficient $\phi\in[0,\bar\phi]$, which obtains when every unit's disturbance is such an autoregression with a common coefficient, or more generally when the units' coefficients are all at most $\bar\phi$ and the units are independent, so that the gap's spectral density is a mixture of autoregressive densities, $\underline\rho=(1-\bar\phi)/(1+\bar\phi)$ and $\overline\rho=(1+\bar\phi)/(1-\bar\phi)$ whatever the gaps between observation times, since the spectral density of the process is bounded by those numbers and the eigenvalues of a principal submatrix of its correlation matrix lie between its extreme eigenvalues.

\begin{proposition}[A conservative calibration under bounded dependence]\label{prop:dependence}
Let Assumption~\ref{ass:bias} hold, let $\eta$ on $\Ptwo\cup\Q$ be conditionally Gaussian with covariance $\omega^2\bar R$ as above, and let $\widehat\sigma_\lambda$ and $\nu_\lambda$ be the scale and degrees of freedom of \eqref{eq:scale}, computed without any whitening. Let $\chi^2_{\nu,\alpha/2}$ denote the $\alpha/2$ quantile of the chi-square distribution with $\nu$ degrees of freedom. Then the interval
\begin{equation}\label{eq:dep-ci}
\Big[\widehat\theta(\lambda)\ \pm\ \big\{B(\lambda)+k^{\mathrm{dep}}_{\alpha}\widehat\sigma_\lambda\big\}\Big],\qquad
k^{\mathrm{dep}}_{\alpha}=z_{1-\alpha/4}\sqrt{\frac{\overline\rho}{\underline\rho}}\sqrt{\frac{\nu_\lambda}{\chi^2_{\nu_\lambda,\alpha/2}}},
\end{equation}
has conditional coverage at least $1-\alpha$ for every admissible law, with or without contamination of the validation periods.
\end{proposition}

The union bound is the price of making no use of the joint law of the estimation error and the scale. When the dependence belongs to a parametric family $\bar R(\phi)$, $\phi\in\Phi$, that law is known up to $\omega$ for every $\phi$, and a sharper critical value is available by computation. Write $\xi(\lambda)$ for the Gaussian part of the estimation error, so that $|\widehat\theta-\theta|\le B(\lambda)+|\xi(\lambda)|$ pathwise, and define
\begin{equation}\label{eq:k-num}
k^{\mathrm{num}}_\alpha=\inf\Big\{k\ge0\ :\ \sup_{\phi\in\Phi}\Prob_\phi\big(|\xi(\lambda)|>k\,\widehat\sigma_\lambda\ \big|\ \F\big)\le\alpha\Big\}.
\end{equation}
The ratio $|\xi(\lambda)|/\widehat\sigma_\lambda$ is a function of the standardised disturbances alone, so its law under $\phi$ depends on neither $\omega$ nor the discrepancy, and it is the ratio of the absolute value of a linear form to the square root of a quadratic form in a Gaussian vector of dimension $n+m$, which Monte Carlo evaluates to any precision in seconds. Since non-coverage of $[\widehat\theta\pm(B+k\widehat\sigma_\lambda)]$ is contained in $\{|\xi|>k\widehat\sigma_\lambda\}$, the interval with $k^{\mathrm{num}}_\alpha$ has conditional coverage at least $1-\alpha$ for every admissible law with $\K=\emptyset$ and every $\phi\in\Phi$. With contamination the scale is inflated by an $\F$-measurable shift, and the Anderson argument of the proof of Proposition~\ref{prop:dependence} applies to the scale alone, not to the joint event, so the numerical calibration is stated for $S=0$ and the union bound covers the rest. For the design of Section~\ref{sec:evidence} with $\nu_\lambda=4$ and $\Phi=[0,0.5]$, $k^{\mathrm{num}}_{0.05}=4.16$, the probability being computed on a grid of eleven values of $\phi$ and increasing in $\phi$ over the grid, against $t_{4,0.975}=2.78$ and the union-bound value of $19.3$. The numerical calibration is the one we recommend when a bound on the persistence can be defended, and the proposition is the statement of what remains true without any computation.

The proposition's calibration is a union bound. With probability at least $1-\alpha/2$ the residual scale is not too small relative to its independent-noise value, and on that event the estimation error, whose variance is at most $\overline\rho$ times its independent-noise value, is covered with probability at least $1-\alpha/2$. The two events are not independent, which is why the bound is a union rather than a product, and the price is large. With $\nu_\lambda=4$ and $\bar\phi=0.5$, $k^{\mathrm{dep}}_{0.05}=19.3$ against $t_{4,0.975}=2.78$, and with $\nu_\lambda=10$ it is $11.8$ against $2.23$. The proposition is therefore a guarantee rather than a recommendation. Its use is to state what finite-sample validity costs when dependence cannot be excluded, and to show that the time-split construction does not depend on independence for its logic, only for its sharpness. When the dependence is known the whitening of Remark~\ref{rem:R} restores the exact Student calibration, and when the excluded periods between the validation block and the post-treatment period are numerous relative to the memory of the process, the correlation between $\xi(\lambda)$ and the residual scale is small and the Student interval is close to valid, as the simulations of Section~\ref{sec:evidence} show under moderate persistence.

Because $k^{\mathrm{num}}_\alpha$ is computed for a stated class, it supports a sensitivity analysis in the persistence bound in the same way that Definition~\ref{def:breakdown} supports one in the curvature bound. For the autoregressive classes $\Phi=[0,\bar\phi]$, the breakdown persistence $\bar\phi^*$ is the smallest $\bar\phi$ on a grid at which the interval with $k^{\mathrm{num}}_\alpha(\bar\phi)$ includes zero. The critical value is computed on the actual time positions of the validation and post-treatment periods, so excluded periods between the blocks enter through the decay of the correlations. The breakdown persistence is the number a reader who doubts independence should ask for, and Section~\ref{sec:abadie} reports it alongside the breakdown curvature.

\section{Simulation}\label{sec:evidence}
The evidence comes in two parts. A controlled Gaussian experiment isolates the coverage calculation of Theorem~\ref{thm:split} after the training decisions have been fixed, and an end-to-end experiment re-estimates the donor weights, calibrates the bounds and selects the anchor in every replication, so that the construction is compared with the conventional one and with four competing methods, in six versions, on the task a practitioner faces.

\subsection{An Illustrative Conditional Experiment}
The first experiment simulates the conditional experiment in Proposition~\ref{prop:decomposition}. There are $n=7$ validation periods and $m=3$ post-treatment periods, with
\begin{equation}\label{eq:simulation-design}
 x_{\Ptwo}=(-7,-6,-5,-4,-3,-2,-1)',\qquad \bar x_{\Q}=1,
\end{equation}
and the affine anchor is
\begin{equation}\label{eq:simulation-anchor}
 \lambda=(-0.10,0.05,0.05,0.10,0.20,0.30,0.40)'.
\end{equation}
The gap variance is $\omega^2=1$. We set $\theta=0$, $c=0$, $d=M=0.05$, and $\ell=L=0.15$. The third validation observation, corresponding to $x=-5$, is the sole potentially contaminated period. Its mean disturbance is $\kappa_3=-S$, with all other $\kappa_t=0$. The contamination bound takes values $S\in\{0,1,4\}$. These choices give
\begin{equation}\label{eq:simulation-boundary}
 h(\lambda)=2.85,\qquad B(\lambda)=0.2925+0.05S,
 \qquad \beta(\lambda)=B(\lambda).
\end{equation}
Thus the discrepancy is at the positive boundary of the scalar bias envelope. The estimator's standard deviation is $\sigma=0.80519149$ and the residual projection has $\nu_\lambda=4$ degrees of freedom.

Each design uses $250{,}000$ independent replications. Within a replication, the validation disturbances are independent standard normals and the post-treatment average disturbance is an independent $N(0,1/3)$ variable. The same simulated disturbances are used across contamination levels and competing intervals. We compare the feasible interval~\eqref{eq:honest-ci}, an unadjusted Student interval $[\widehat\theta\pm t_{4,0.975}\widehat\sigma]$, and the known-scale interval~\eqref{eq:gaussian-ci}. The nominal coverage probability is $0.95$ throughout.

\begin{table}[htbp]
\centering\small
\begin{threeparttable}
\caption{Conditional coverage in the Gaussian experiment}\label{tab:simulation}
\begin{tabular}{@{}rrrrrr@{}}
\toprule
\multicolumn{6}{@{}l}{\emph{A. Contamination of one validation period, independent noise}}\\
$S$ & $B$ & Feasible & Unadjusted & Known scale & Mean length\\
\midrule
0 & 0.2925 & 0.9696 & 0.9406 & 0.9500 & 4.785 \\
1 & 0.3425 & 0.9785 & 0.9524 & 0.9501 & 5.290 \\
4 & 0.4925 & 0.9996 & 0.9986 & 0.9504 & 9.843 \\
\addlinespace
\multicolumn{6}{@{}l}{\emph{B. First-order autoregressive noise, no contamination}}\\
$\phi$ & Student & Numerical $k$ & Union bound $k$ & \multicolumn{2}{l}{Mean lengths} \\
\midrule
0.0 & 0.9692 & 0.9914 & 1.0000 & 4.79 / 6.89 / 29.85 \\
0.3 & 0.9398 & 0.9816 & 1.0000 & 4.25 / 6.09 / 26.12 \\
0.5 & 0.9116 & 0.9696 & 0.9998 & 3.75 / 5.33 / 22.62 \\
\bottomrule
\end{tabular}
\begin{tablenotes}[flushleft]\footnotesize
\item \emph{Notes:} Panel A, 250,000 replications per row with the design of Section~\ref{sec:evidence}, $\nu_\lambda=4$, $\sigma=0.8052$, discrepancy at the positive boundary of the bias envelope. Feasible is the interval \eqref{eq:honest-ci}, unadjusted omits the bias allowance, known scale is \eqref{eq:gaussian-ci}. Mean length refers to the feasible interval. Monte Carlo standard errors of coverage are below 0.0005. Panel B, 50,000 replications with $S=0$, the same anchor and bounds, and Gaussian noise that follows a first-order autoregression with coefficient $\phi$ across the ten consecutive validation and post-treatment periods. Student uses $t_{4,0.975}=2.78$. Numerical $k$ is the critical value \eqref{eq:k-num} calibrated over the autoregressive class with $\bar\phi=0.5$, $k^{\mathrm{num}}_{0.05}=4.16$, and union bound $k$ is \eqref{eq:dep-ci}, $k^{\mathrm{dep}}_{0.05}=19.3$. Lengths are reported in that order.
\end{tablenotes}
\end{threeparttable}
\end{table}

Panel A of Table~\ref{tab:simulation} illustrates the two sources of conservatism in the feasible interval. With $S=0$, its coverage is $0.970$, while the interval omitting the bias allowance covers with probability $0.941$. As contamination increases, the feasible interval widens because both the bias allowance and the residual scale increase. At $S=4$, its average length is about twice the uncontaminated value. The unadjusted interval also becomes conservative as the residual scale is inflated. Such overcoverage is specific to the contaminated designs and does not provide a uniform guarantee when the discrepancy bound is omitted. The known-scale interval remains close to its exact boundary coverage of $0.95$ in each row. Panel B replaces the independent noise by a first-order autoregression over the ten consecutive validation and post-treatment periods. The Student interval loses coverage as persistence rises, to $0.940$ at $\phi=0.3$ and $0.912$ at $\phi=0.5$, because the residual scale understates the variance of the estimator when the disturbances are positively correlated. The numerically calibrated critical value \eqref{eq:k-num} with $\bar\phi=0.5$ covers in every row, at 0.970 where the Student interval covers 0.912, and lengthens the interval by about two fifths. The union bound of Proposition~\ref{prop:dependence} also covers in every row, at six times the Student length, which is the price of a guarantee that uses nothing but the eigenvalue bounds.

\subsection{An End-to-End Experiment with Estimated Weights}\label{sec:endtoend}

The second experiment re-estimates everything. Each replication draws fourteen donors and one treated unit over eleven pre-treatment periods, split into a training block of five and a validation block of six, one excluded period and ten post-treatment periods. The treated unit's level lies 0.8 below the lowest donor's, so the convex-hull condition fails by construction, and donors are as noisy as the treated unit except in design D. Donor weights are fitted by de-meaned synthetic control on the training block in every replication. Design A draws the treated unit's growth like the donors', design C sets it 1.5 points a year below the slowest donor's, design D gives the treated unit persistent noise with autoregressive coefficient 0.85, which the maintained model excludes, and design B2 places an own shock inside the validation block that the drift bound does not cover. Appendix~\ref{app:sim} gives the details. Table~\ref{tab:sim} reports coverage and median length at a nominal level of 95 percent and Table~\ref{tab:simpower} in the appendix the rejection rate of a zero effect. The competitors are the placebo set of \citet{abadie2010} built with the same split weights, the conformal set of \citet{chernozhukov2021} in two versions, the $t$-test of \citet{chernozhukov2021t} with three blocks and simplex or de-meaned weights, and the prediction interval of \citet{cattaneo2021} for the average effect over the post-treatment periods.

\begin{table}[tp]
\centering\scriptsize\setlength{\tabcolsep}{3pt}\renewcommand{\arraystretch}{0.92}
\begin{threeparttable}
\caption{Simulations with estimated weights, nominal level 95 percent}\label{tab:sim}
\begin{tabular}{@{}p{5.7cm}cccc@{\hspace{10pt}}cccc@{}}
\toprule
 & \multicolumn{4}{c}{Coverage} & \multicolumn{4}{c}{Median length} \\
\cmidrule(lr){2-5}\cmidrule(l){6-9}
 & A & C & D & B2 & A & C & D & B2 \\
\midrule
Conventional, $M$ known & 0.82 & 0.99 & 0.61 & 0.09 & 0.06 & 0.43 & 0.22 & 0.53 \\
Conventional, $M=0$ & 0.29 & 0.00 & 0.21 & 0.07 & 0.03 & 0.09 & 0.07 & 0.13 \\
Conventional, $M$ placebo & 0.95 & 0.30 & 0.73 & 0.21 & 0.23 & 0.29 & 0.28 & 0.33 \\
Conventional, placebo scale, $M=0$ & 0.92 & 0.01 & 0.50 & 0.06 & 0.17 & 0.17 & 0.17 & 0.17 \\
Conventional, placebo scale, $M$ known & 1.00 & 1.00 & 0.81 & 0.28 & 0.20 & 0.50 & 0.32 & 0.58 \\
Split weights, validation noise, normal cv & 0.92 & 0.98 & 0.60 & 0.26 & 0.13 & 0.39 & 0.25 & 0.26 \\
Split, uniform, $M$ known$^{\dagger}$ & 0.97 & 0.97 & 0.68 & 0.52 & 0.15 & 0.41 & 0.28 & 0.40 \\
Split, uniform, $M=0$ & 0.38 & 0.00 & 0.30 & 0.37 & 0.06 & 0.07 & 0.11 & 0.31 \\
Split, uniform, $M$ placebo & 0.96 & 0.57 & 0.81 & 0.94 & 0.35 & 0.36 & 0.41 & 0.60 \\
Split, optimal, $M$ placebo & 0.98 & 0.94 & 0.83 & 0.99 & 0.27 & 0.30 & 0.43 & 1.08 \\
Split, trend-balancing$^{\dagger}$ & 0.95 & 0.95 & 0.77 & 0.99 & 0.27 & 0.31 & 0.48 & 1.31 \\
Theorem 3, uniform$^{\ddagger}$ & 0.99 & 0.85 & 0.91 & 1.00 & 0.46 & 0.48 & 0.59 & 1.07 \\
Theorem 3, data-chosen anchor$^{\ddagger}$ & 0.97 & 0.97 & 0.89 & 1.00 & 0.33 & 0.37 & 0.52 & 1.03 \\
Split, uniform, $M,S$ known$^{\dagger}$ &  &  &  & 1.00 &  &  &  & 0.90 \\
Placebo set, split & 0.93 & 0.31 & 0.65 & 0.39 & 0.29 & 0.29 & 0.29 & 0.29 \\
Conformal (CWZ), cyclic & 0.94 & 0.41 & 0.97 & 0.96 & 0.85 & 0.95 & 0.90 & 1.00 \\
Conformal (CWZ), trend model & 1.00 & 1.00 & 1.00 & 1.00 & 0.68 & 0.70 & 1.40 & 1.87 \\
$t$-test (CWZ), $K=3$, simplex & 0.42 & 0.07 & 0.64 & 0.92 & 0.28 & 0.78 & 0.41 & 1.00 \\
$t$-test (CWZ), $K=3$, de-meaned & 0.89 & 0.36 & 0.81 & 0.55 & 0.18 & 0.35 & 0.49 & 0.65 \\
Prediction interval (CFT), average & 0.99 & 1.00 & 0.85 & 0.40 & 0.24 & 0.58 & 0.49 & 0.78 \\
\bottomrule
\end{tabular}
\begin{tablenotes}[flushleft]\footnotesize
\item \emph{Notes:} 1,000 replications per design, $J=14$ donors, eleven pre-treatment periods split into a training block of five and a validation block of six, one excluded period, ten post-treatment periods, true effect $-0.30$. Design A, the treated unit grows like the donors. C, it grows 1.5 points a year below the slowest donor. D, its noise is a first-order autoregression with coefficient 0.85. B2, an own shock in the validation block. Weights by de-meaned synthetic control on the training block (split rows) or on all pre-treatment periods (conventional rows). Rows marked $M$ known use the realised drift of the fitted weights, and rows marked $M$ placebo use the largest placebo error divided by $|h_U|$, and Theorem 3 rows use the bound \eqref{eq:Mhat} with the simultaneous critical value $k^{\mathrm{sim}}_{0.05}=3.48$, guarantee $1-0.05-1/15=0.883$. The optimal anchor is selected with the training-block scale of Section~\ref{sec:anchor}. Conventional rows use the in-sample residual scale and the normal critical value, or the root mean square of the full-window placebo estimates (placebo scale). The $t$-test of \citet{chernozhukov2021t} uses $K=3$ blocks of $r=3$ pre-treatment periods. The prediction interval of \citet{cattaneo2021} is for the average effect over the post-treatment periods, computed with the authors\textquoteright{} scpi package, simplex weights with an intercept, Gaussian out-of-sample bounds and 200 simulations. Rows marked $\dagger$ carry the finite-sample guarantee of Theorem~\ref{thm:split} when the maintained model holds and rows marked $\ddagger$ the guarantee of Theorem~\ref{thm:exch}, and the remaining rows carry none. No design satisfies the maintained model exactly, because the common cycle of Appendix~\ref{app:sim} adds an imbalance that is neither drift nor independent noise and that the bound $M$ known does not cover. The row on split weights with validation noise fits the weights on the training block and uses the uniform anchor, the root mean square of the validation residuals around it and the normal critical value. Monte Carlo standard errors of coverage are at most 0.016. Rejection rates of a zero effect are in Table~\ref{tab:simpower}.
\end{tablenotes}
\end{threeparttable}
\end{table}

Four findings stand out. First, the conventional construction fails even when the bias bound is known. It covers 82 percent of the time in design A, where the treated unit's growth is drawn like the donors', and 61 percent under persistent noise. It covers 99 percent when growth lies outside the hull, not because it is valid but because the in-sample residuals then contain the drift and overstate the noise. With the bound set to zero it covers 29, 0 and 21 percent in the three designs without a shock. Fitting the weights on the training block but estimating the noise from the validation residuals around the anchor, without the projection or the Student quantile, covers 92 percent in design A, so the weights account for about three quarters of the deficit there and the scale estimate and critical value for the rest. Replacing the in-sample scale by the root mean square of the full-window placebo estimates gives 92 percent in design A and 1 percent in design C with no drift allowance and 100 percent in both with the known one, because the placebos' own drift then enters the scale on top of the bias bound.

Second, the time-split interval covers at or above its level in every design in which its bounds hold and the disturbances are independent, and it is only as honest as its bounds. With a known drift bound the uniform anchor covers 97 percent in designs A and C. With the bound set to zero it covers 38, 0 and 30 percent in designs A, C and D. The trend-balancing anchor, which needs no drift bound, covers 95 percent in designs A and C at a length of 0.27 and 0.31, and it is the only split construction with a bound fixed in advance that covers under the validation-block shock without being told its size or location, at 99 percent with a length of 1.3 and no power. The placebo-calibrated constructions cover there too, because the shock inflates the placebo errors from which their bound is computed. Under persistent noise every split construction with a drift allowance undercovers, by between 4 and 27 percentage points, which is the cost of the independence assumption and the reason Section~\ref{sec:dependence} exists.

Third, the placebo-calibrated bounds behave as Theorem~\ref{thm:exch} predicts, and Table~\ref{tab:cond} shows why. The raw placebo bound with the uniform anchor covers 96 percent in design A and 57 percent in design C, where the treated unit's growth lies outside the donors' range. On the event $E_1$ that the treated unit's uniform-anchor error is no larger than the largest placebo error, the uniform anchor covers in every replication of every design, and off it the coverage falls to between 37 and 49 percent in the designs without a shock. The event has probability 0.93 in design A, as exchangeability with fourteen donors requires, and 0.31 in design C, so the failure in design C is a failure of exchangeability and not of the time split. The bound of Theorem~\ref{thm:exch}, which adds the treated unit's own noise and uses the simultaneous critical value, covers 99 percent in design A and 85 percent in design C with the uniform anchor, against a guarantee of 0.883. With the anchor chosen from the data to minimise the length it covers 97 percent in designs A and C at lengths of 0.33 and 0.37, a fifth longer than the trend-balancing interval, because the data move the anchor toward trend balance whenever the placebo errors are large, and its coverage off $E_1$ is 97 percent in both designs. The optimal anchor with the raw placebo bound, selected with the training-block scale, behaves the same way for the same reason, covering 98 and 94 percent.

\begin{table}[tp]
\centering\scriptsize\setlength{\tabcolsep}{3.5pt}
\begin{threeparttable}
\caption{Coverage conditional on placebo exchangeability holding}\label{tab:cond}
\begin{tabular}{@{}p{4.8cm}cccccccc@{}}
\toprule
 & \multicolumn{2}{c}{A} & \multicolumn{2}{c}{C} & \multicolumn{2}{c}{D} & \multicolumn{2}{c}{B2} \\
\cmidrule(lr){2-3}\cmidrule(lr){4-5}\cmidrule(lr){6-7}\cmidrule(l){8-9}
 & $E_1$ & not $E_1$ & $E_1$ & not $E_1$ & $E_1$ & not $E_1$ & $E_1$ & not $E_1$ \\
\midrule
Probability of the event & 0.93 & 0.07 & 0.31 & 0.69 & 0.65 & 0.35 & 0.39 & 0.61 \\
Probability that the placebo drift bound holds & \multicolumn{2}{c}{0.93} & \multicolumn{2}{c}{0.30} & \multicolumn{2}{c}{0.78} & \multicolumn{2}{c}{0.92} \\
Probability that the bound of Theorem 3 holds & \multicolumn{2}{c}{0.97} & \multicolumn{2}{c}{0.63} & \multicolumn{2}{c}{0.92} & \multicolumn{2}{c}{1.00} \\
\addlinespace
Placebo set, split & 1.00 & 0.00 & 1.00 & 0.00 & 1.00 & 0.00 & 1.00 & 0.00 \\
Split, uniform, $M$ placebo & 1.00 & 0.49 & 1.00 & 0.37 & 1.00 & 0.44 & 1.00 & 0.89 \\
Split, optimal, $M$ placebo & 0.98 & 0.96 & 0.98 & 0.93 & 0.85 & 0.80 & 1.00 & 0.98 \\
Theorem 3, uniform & 1.00 & 0.83 & 1.00 & 0.79 & 1.00 & 0.75 & 1.00 & 1.00 \\
Theorem 3, data-chosen anchor & 0.97 & 0.97 & 0.96 & 0.97 & 0.92 & 0.84 & 1.00 & 1.00 \\
Conventional, $M$ placebo & 0.98 & 0.58 & 0.51 & 0.20 & 0.86 & 0.49 & 0.39 & 0.09 \\
Prediction interval (CFT), average & 0.99 & 0.99 & 1.00 & 1.00 & 0.91 & 0.75 & 0.53 & 0.31 \\
\bottomrule
\end{tabular}
\begin{tablenotes}[flushleft]\footnotesize
\item \emph{Notes:} $E_1$ is the event $|\Pi_1|\le\max_{j\ge2}|\Pi_j|$ of the proof of Theorem~\ref{thm:exch}, on which the treated unit\textquoteright{}s uniform-anchor error is no larger than the largest placebo error. Its probability is at least $1-1/15=0.933$ under Assumption~\ref{ass:exch}. The placebo drift bound holds when the largest placebo error divided by $|h_U|$, and the bound of Theorem 3 when $\widehat M$ of \eqref{eq:Mhat}, is at least the realised drift of the treated unit\textquoteright{}s weights. Coverage among the replications in which the event holds and in which it fails. Designs and rows as in Table~\ref{tab:sim}.
\end{tablenotes}
\end{threeparttable}
\end{table}

Fourth, the competitors differ sharply. The prediction interval of \citet{cattaneo2021} for the average effect is the strongest. It covers 99 and 100 percent in designs A and C, 85 percent under persistent noise and 40 percent under the validation-block shock, at lengths of 0.24, 0.58, 0.49 and 0.78, and it is shorter than every feasible split construction that covers in design A. The split constructions improve on it in design C, where the trend-balancing anchor and the data-chosen anchor of Theorem~\ref{thm:exch} cover at lengths of 0.31 and 0.37 against 0.58, and under the shock, where they cover and it does not. The $t$-test of \citet{chernozhukov2021t} undercovers in this panel. With simplex weights it covers 42 percent in design A and 7 percent in design C, because the treated unit lies outside the donors' hull, so that simplex weights without an intercept leave a gap that drifts, and a drifting gap biases every fold estimate in the same direction by an amount that the dispersion of the fold estimates does not measure, and with de-meaned weights it covers 89 and 36 percent. Moving the three blocks to the end of the pre-treatment window, which \citet{chernozhukov2021t} allow, does not help, since the de-meaned version then covers 78 percent in design A and 44 percent in design C at shorter lengths. Its validity rests on stationarity and on many pre-treatment and post-treatment periods, neither of which the design provides, so the result is a statement about short panels with drift rather than about the test in the setting for which it was built. The conformal set with cyclic permutations covers or nearly covers in designs A, D and B2, at 94 to 97 percent, but not in C, at lengths near one, and the version with a fitted trend and i.i.d.\ permutations covers everywhere at lengths of 0.68 to 1.87, with power of 42 and 39 percent in designs A and C and almost none in D and B2. The placebo set with the split weights covers 93 percent in design A and 31 percent in design C at a length of 0.29.

The split has a cost in the precision of the point estimate as well as in length. The root mean square error of the trend-balancing estimate is 0.061 in design A against 0.048 for the conventional estimate, and 0.064 in design C against 0.179, where the conventional estimate inherits the drift. The reader should take from the table that the time split buys finite-sample validity for an explicit class, that the class must be right for the validity to be worth anything, that a placebo calibration is worth exactly as much as the exchangeability on which it rests, and that a prediction interval which bounds the drift by its own pre-treatment size is the appropriate alternative when no bound can be defended.

\subsection{Curvature Designs}\label{sec:curvsim}

The designs above have linear discrepancies, so the curvature class is never tested by them. Table~\ref{tab:curv} adds a quadratic term $\gamma_i(x_t-x_0)^2$ to every unit's untreated path in design A, with $x_0$ the last validation period and $\gamma_i$ uniform on $[-0.004,0.004]$ across donors, so that second derivatives of up to 0.008 log points per squared period are present in the donor pool. In design E1 the treated unit's curvature is drawn from the same range, and in design E2 it is $-0.012$ per squared period, outside it. The slope of the discrepancy is left unrestricted throughout.

\begin{table}[tp]
\centering\scriptsize\setlength{\tabcolsep}{4pt}
\begin{threeparttable}
\caption{Curvature designs, nominal level 95 percent}\label{tab:curv}
\begin{tabular}{@{}p{4.8cm}cccc@{\hspace{8pt}}cccc@{}}
\toprule
 & \multicolumn{4}{c}{E1 curvature inside donors\textquoteright{} range} & \multicolumn{4}{c}{E2 curvature outside donors\textquoteright{} range} \\
\cmidrule(lr){2-5}\cmidrule(l){6-9}
 & Coverage & \shortstack{Coverage,\\bound holds} & Length & Power & Coverage & \shortstack{Coverage,\\bound holds} & Length & Power \\
\midrule
Split, trend-balancing, linear class & 0.77 & 0.80 & 0.28 & 0.89 & 0.23 & 0.34 & 0.33 & 1.00 \\
Split, trend-balancing, $C$ placebo & 0.99 & 1.00 & 0.63 & 0.45 & 0.80 & 1.00 & 0.68 & 0.93 \\
Split, trend-balancing, $C$ known & 0.97 & 0.97 & 0.42 & 0.76 & 0.97 & 0.99 & 0.81 & 0.89 \\
$t$-test (CWZ), $K=3$, simplex & 0.77 & 0.79 & 0.71 & 0.44 & 0.88 & 0.89 & 1.17 & 0.30 \\
$t$-test (CWZ), $K=3$, de-meaned & 0.67 & 0.70 & 0.26 & 0.84 & 0.73 & 0.69 & 0.59 & 0.73 \\
Prediction interval (CFT), average & 0.78 & 0.80 & 0.34 & 0.81 & 0.68 & 0.71 & 0.73 & 0.64 \\
\bottomrule
\end{tabular}
\begin{tablenotes}[flushleft]\footnotesize
\item \emph{Notes:} 1,000 replications per design, otherwise design A of Table~\ref{tab:sim} with a quadratic term $\gamma_i(x_t-x_0)^2$ added to every unit\textquoteright{}s untreated path, $x_0$ the last validation period, $\gamma_i$ uniform on $[-0.004,0.004]$ for the donors and for the treated unit in E1, and $\gamma_1=-0.006$ in E2. The slope of the discrepancy is unrestricted (Assumption~\ref{ass:curv}). $C$ placebo is the largest placebo curvature, $C$ known the realised curvature $2|\gamma_1-w^{\prime}\gamma|$ of the fitted weights. The placebo bound is at least the realised curvature in 93 percent of replications in E1 and 16 percent in E2, and the second coverage column conditions on that event. The curvature-optimal anchor, selected with the training-block scale and compared with the trend-balancing anchor at their own degrees of freedom, coincides with the trend-balancing anchor in this design and is not shown. Length is the median, power the rejection rate of a zero effect.
\end{tablenotes}
\end{threeparttable}
\end{table}

The results are what the two classes imply. Ignoring curvature is costly. The trend-balancing interval under the linear class covers 77 percent in E1 and 23 percent in E2. The placebo curvature bound repairs this exactly when it is valid. It is at least the realised curvature in 93 percent of replications in E1, and the trend-balancing interval at that bound covers 99 percent, at a length of 0.63 against 0.28 under the linear class, which is the price of admitting curvature of the size the donors display. In E2 the placebo bound holds in 16 percent of replications and the same interval covers 80 percent, against 97 percent with the realised curvature, and it covers in every replication in which the bound holds. This is the same failure as that of the placebo drift bound in design C and has the same cause. With six validation periods the curvature-optimal anchor never beats trend balance once each candidate is charged its own degrees of freedom, because leaving the span of the intercept and the trend costs one of four degrees of freedom, so in this design it selects the trend-balancing anchor. A curvature bound calibrated from the most curved placebo is therefore a defensible restriction when the treated unit's path bends no more than the donors' do, and no restriction at all otherwise, and the breakdown curvature of Section~\ref{sec:abadie} should be read with that in mind. Curvature also costs the competitors. The prediction interval of \citet{cattaneo2021} covers 78 percent in E1 and 68 percent in E2, because a bound on the post-treatment error calibrated from pre-treatment residuals does not anticipate a slope that changes, and the $t$-test of \citet{chernozhukov2021t} covers between 67 and 88 percent depending on the weights.

\section{Applications}\label{sec:application}\label{sec:abadie}

\begin{figure}[tp]
\centering
\includegraphics[width=\textwidth]{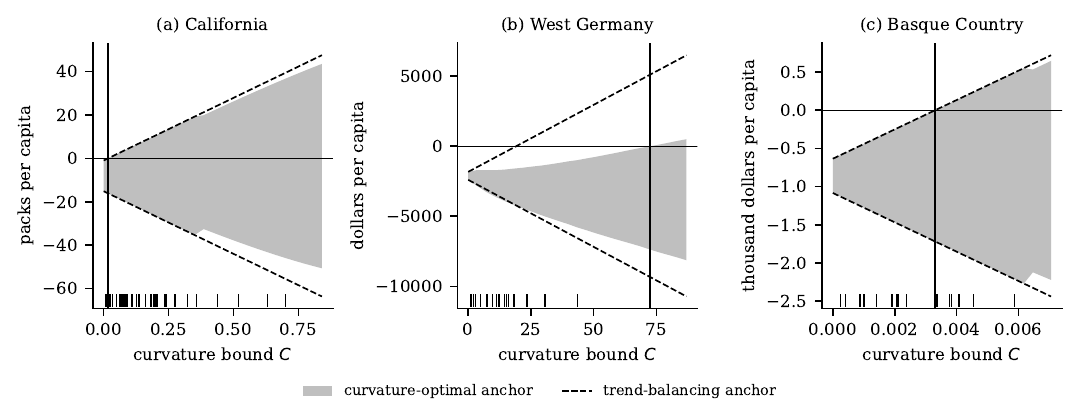}
\caption{Breakdown curvature in the three classic applications}\label{fig:abadie}
\fignote{De-meaned weights. Each panel plots the 95 percent interval for the average effect against the curvature bound $C$ of Assumption~\ref{ass:curv}, with the slope of the discrepancy unrestricted. The shaded band is the interval with the curvature-optimal anchor and the dashed lines the interval with the trend-balancing anchor. The vertical line marks the breakdown curvature $C^*$ of the curvature-optimal interval, and the ticks along the bottom mark the curvatures of the placebo gaps.}
\end{figure}

The applications are the effect of California's Proposition 99 on cigarette sales per capita \citep{abadie2010}, with 38 donor states, a pre-treatment window of 1970-1988 and a post-treatment window of 1989--2000, the effect of reunification on West German GDP per capita \citep{abadie2015}, with 16 OECD donors, 1960-1990 and 1991-2003, and the effect of terrorism on Basque GDP per capita \citep{abadie2003}, with 16 Spanish regions, 1955--1969 and 1970-1997. Outcomes are in the levels the original studies use. Each pre-treatment window is split at its midpoint, so that the training blocks are 1970-1979, 1960-1975 and 1955-1961 and the validation blocks 1980-1988, 1976-1990 and 1962-1969. Weights are fitted on the training block by de-meaned synthetic control and, in Table~\ref{tab:abadie-sc} of the appendix, by synthetic control without an intercept. The covariates of the original studies are not used, so these are re-analyses of the outcome data and not replications of the original specifications. Table~\ref{tab:abadie} reports the results and Figure~\ref{fig:abadie} draws the breakdown analysis.

\begin{table}[tp]
\centering\scriptsize\setlength{\tabcolsep}{3pt}\renewcommand{\arraystretch}{0.95}
\begin{threeparttable}
\caption{Three classic applications, de-meaned weights}\label{tab:abadie}
\begin{tabular}{@{}l lll@{}}
\toprule
 & California & West Germany & Basque Country \\
\midrule
Donors $J$ & 38 & 16 & 16 \\
Weights, training block & 1970--1979 & 1960--1975 & 1955--1961 \\
Anchor, validation block & 1980--1988 & 1976--1990 & 1962--1969 \\
Largest weights & Nevada 0.32, Connecticut 0.28 & Netherlands 0.38, Greece 0.21 & Madrid 0.41, Murcia 0.37 \\
Placebo curvature $C_{\mathrm{don}}$ & 0.700 & 43.5 & 0.0059 \\
Breakdown $C^*$, optimal anchor & 0.017 (34 of 38 above) & 72.5 (0 of 16 above) & 0.0033 (6 of 16 above) \\
Breakdown $C^*_{\mathrm{tr}}$, trend-balancing anchor & 0.017 (34 of 38 above) & 19.1 (4 of 16 above) & 0.0033 (6 of 16 above) \\
Breakdown persistence $\bar\phi^*$, linear class & 0.2 & $>0.9$ & $>0.9$ \\
\addlinespace
\multicolumn{4}{@{}l}{\emph{Curvature class, $C=C_{\mathrm{don}}$}} \\
\quad Curvature-optimal anchor & $-$4.6 [$-$45.9, 36.8] & $-$3,392 [$-$5,793, $-$991] & $-$0.86 [$-$2.21, 0.49] \\
\quad Trend-balancing anchor & $-$8.1 [$-$55.6, 39.5] & $-$2,112 [$-$6,559, 2,335] & $-$0.86 [$-$2.21, 0.49] \\
\multicolumn{4}{@{}l}{\emph{Linear class ($C=0$)}} \\
\quad Trend-balancing anchor & $-$8.1 [$-$15.1, $-$1.0] & $-$2,112 [$-$2,398, $-$1,826] & $-$0.86 [$-$1.08, $-$0.63] \\
\quad Trend-balancing anchor, AR(1), $\bar\phi=0.5$ & $-$8.1 [$-$19.7, 3.6] & $-$2,112 [$-$2,599, $-$1,626] & $-$0.86 [$-$1.23, $-$0.49] \\
\quad Theorem 3, data-chosen anchor & $-$8.1 [$-$16.6, 0.5] & $-$2,112 [$-$2,451, $-$1,774] & $-$0.86 [$-$1.14, $-$0.58] \\
\quad Uniform anchor, $M=0$ & $-$15.0 [$-$17.2, $-$12.8] & 629 [511, 747] & $-$0.15 [$-$0.19, $-$0.12] \\
\multicolumn{4}{@{}l}{\emph{Other constructions}} \\
\quad Conventional, in-sample noise & $-$11.1 [$-$11.8, $-$10.4] & $-$1,728 [$-$1,771, $-$1,684] & $-$0.94 [$-$0.98, $-$0.90] \\
\quad Conventional, placebo scale & $-$11.1 [$-$30.8, 8.6] & $-$1,728 [$-$5,096, 1,641] & $-$0.94 [$-$1.75, $-$0.13] \\
\quad $t$-test (CWZ), $K=3$ & $-$14.8 [$-$45.2, 15.6] & $-$1,866 [$-$3,164, $-$568] & $-$0.77 [$-$1.46, $-$0.08] \\
\quad Prediction interval (CFT), average & [$-$16.6, 3.4] & [$-$2,940, $-$469] & [$-$1.93, $-$0.59] \\
\quad Conformal (CWZ), cyclic & [$-$99.1, 90.5] & empty & [$-$3.69, 1.00] \\
\bottomrule
\end{tabular}
\begin{tablenotes}[flushleft]\scriptsize
\item \emph{Notes:} Average effect over the post-treatment years (1989--2000, 1991--2003, 1970--1997), estimate followed by the 95 percent interval, in packs per capita, US dollars per capita and thousand 1986 US dollars per capita. Weights are simplex weights with an intercept (de-meaned synthetic control). Curvature rows use Assumption~\ref{ass:curv} with the slope unrestricted. $C_{\mathrm{don}}$ is the largest placebo curvature, twice the quadratic coefficient of a quadratic fitted to a placebo gap over the validation and post-treatment years. The curvature-optimal anchor minimises the Student length over trend-balanced anchors with the training-block scale. $C^*$ and $C^*_{\mathrm{tr}}$ are breakdown curvatures (Definition~\ref{def:breakdown}), with the number of placebos whose curvature exceeds them. $\bar\phi^*$ is the breakdown persistence of the trend-balancing interval on the grid $0,0.1,\dots,0.9$, with 0 meaning that the interval includes zero under independence. The AR(1) row replaces the Student quantile by $k^{\mathrm{num}}_{0.05}$ of \eqref{eq:k-num} over $\phi\in[0,0.5]$, which assumes that the dependence does not reach back into the training block. The Theorem 3 row uses the bound \eqref{eq:Mhat} with $k^{\mathrm{sim}}_{0.05}$, guarantee $1-0.05-1/(J+1)$. Conventional rows fit the weights on all pre-treatment years, anchor uniformly on them and use the normal critical value with the in-sample residual scale or the root mean square of the full-window placebo estimates. The $t$-test of \citet{chernozhukov2021t} uses three blocks of the pre-treatment window. The prediction interval of \citet{cattaneo2021} is for the average effect, from the authors\textquoteright{} scpi package with simplex weights, an intercept, Gaussian out-of-sample bounds and 1,000 simulations. The conformal set of \citet{chernozhukov2021} is empty for West Germany.
\end{tablenotes}
\end{threeparttable}
\end{table}

\subsection{How Much Curvature Each Result Survives?}

The curvature class is the natural place to start, because it is the weakest restriction under which the construction delivers an interval with the slope left free. In West Germany the interval with the curvature-optimal anchor excludes zero up to $C^*=72.5$ dollars per squared year, and none of the sixteen placebo gaps bends that much, the largest placebo curvature being $C_{\mathrm{don}}=43.5$. At $C_{\mathrm{don}}$ the interval is $[-5{,}793,-991]$ dollars per capita, and with simplex weights the breakdown curvature is 76.8, again above every placebo. In the Basque Country the breakdown curvature is 0.0033 thousand dollars per squared year with de-meaned weights, which six of the sixteen placebos exceed, and 0.0070 with simplex weights, which none exceeds. In California it is 0.017 packs per squared year with de-meaned weights, which thirty-four of the thirty-eight placebos exceed, and zero with simplex weights, whose interval includes zero even under linear drift. On the grid of Figure~\ref{fig:abadie} the set of curvature bounds at which zero is excluded is an interval in every case, so $C^*$ separates the bounds under which the sign is established from those under which it is not.

The breakdown curvature can be read in the units of the outcome. For the trend-balancing anchor, $C^*_{\mathrm{tr}}\kappa_C(\lambda_{\mathrm{tr}})$ is by \eqref{eq:Cstar} the distance of the linear-class interval from zero, 1.0 packs, 1,826 dollars and 0.63 thousand dollars, which is the largest curvature-induced bias in the average effect that the conclusion tolerates. The curvature-optimal anchor is selected with a scale computed from the donors' training-block prediction errors, and the comparison with the trend-balancing anchor at their own degrees of freedom is also made at that scale, so the selected anchor is training-measurable for each fixed $C$. Evaluated at $C_{\mathrm{don}}$, which depends on the donors' later outcomes, it is a sensitivity analysis and not a guarantee, as are all rows at $C_{\mathrm{don}}$. In California and the Basque Country with de-meaned weights the selected anchor is the trend-balancing one. In West Germany it moves weight to the end of the validation block, with weights of $-0.30$, $0.80$ and $2.47$ on 1988, 1989 and 1990, which cuts its curvature exposure and nearly quadruples the breakdown curvature relative to the trend-balancing anchor, whose $C^*_{\mathrm{tr}}$ is 19.1 dollars and is exceeded by four placebos.

That concentration on 1990 is a reason for caution, since the monetary union of July 1990 may have moved West German output before reunification in October. With 1990 removed from the validation block and treated as an excluded period, the trend-balancing interval is $[-2{,}035,-1{,}741]$ and the breakdown curvature of the optimal anchor falls to 33.8 dollars, which one of the sixteen placebos exceeds, and the interval at $C_{\mathrm{don}}$ includes zero. With simplex weights it falls to 21.7, which five placebos exceed. The reunification result is therefore robust to the curvature of every placebo only when 1990 is used as a normal year, and to the curvature of most placebos otherwise. The same check for California, with 1988 removed, gives a trend-balancing interval of $[-19.0,-5.3]$ and a breakdown curvature of 0.083, which twenty-three placebos exceed, so the California result remains fragile.

\subsection{What Linear Extrapolation Adds?}

At $C=0$ the class is linear and the precision of the intervals rests on extrapolating the slope of the gap at the end of the validation block. With the trend-balancing anchor the interval is the classical prediction interval for the mean of $m$ new observations from a line fitted to the validation-block gap (Remark~\ref{rem:pi}). The intervals are $[-15.1,-1.0]$ packs, $[-2{,}398,-1{,}826]$ dollars and $[-1.08,-0.63]$ thousand dollars, between a fifth and seven tenths of the length of the prediction intervals of \citet{cattaneo2021} for the average effect. The weights fitted on the training block drift away from the treated unit once they are evaluated on the validation block, since the root mean square of the de-meaned gap rises from 0.34 to 2.5 packs, from 17 to 857 dollars and from 0.025 to 0.094 thousand dollars between the blocks, so the uniform anchor with no drift allowance is precise and wrong. In West Germany it gives a positive effect of 629 dollars with an interval of $[511,747]$, because the gap was rising through the validation block and the uniform anchor attributes the whole rise to reunification.

The placebo-calibrated drift bound of Theorem~\ref{thm:exch} lets the data choose the anchor. In all three applications the length-minimising member of the family is the trend-balancing anchor, because the placebo errors are large relative to the noise, so the theorem adds no precision here. What it adds is a statement of when the common practice of bounding the drift by the placebos is valid, and the intervals with the simultaneous critical value, $[-16.6,0.5]$, $[-2{,}451,-1{,}774]$ and $[-1.14,-0.58]$, carry guarantees of $0.924$, $0.891$ and $0.891$ under placebo exchangeability.

The split date matters less for West Germany and the Basque Country than for California. With the training block extended to two thirds of the window, so that the validation blocks are 1983-1988, 1981-1990 and 1965-1969, the trend-balancing intervals are $[-14.2,14.0]$, $[-2{,}478,-1{,}505]$ and $[-1.16,-0.86]$, and the breakdown curvatures of the optimal anchor are 0, 53.8 and 0.0051, exceeded by thirty-eight, one and two placebos. The shorter California validation block contains the years in which its gap turns, and its effect disappears under linear drift.

\subsection{Serial Dependence}

Over the autoregressive class with $\bar\phi=0.5$ the trend-balancing intervals become $[-19.7,3.6]$, $[-2{,}599,-1{,}626]$ and $[-1.23,-0.49]$. The breakdown persistence is $\bar\phi^*=0.2$ in California, and on the grid up to 0.9 the interval never includes zero in West Germany or the Basque Country. These numbers need a caveat. The calibration of Section~\ref{sec:dependence} requires the gap disturbances of the validation and post-treatment periods to have mean zero given the training block, which fails for a persistent process whose memory reaches back across the split, and here the blocks are adjacent. The breakdown persistence therefore measures robustness to dependence within the validation and post-treatment periods only. An excluded buffer between the training and validation blocks would make the restriction a better approximation, at the price of validation periods that these panels cannot spare.

\subsection{Competitors}

The conventional in-sample intervals are implausibly narrow, $\pm43$ dollars for West Germany with 31 pre-treatment years and 16 donors, because the in-sample residuals contain neither the drift nor the serial correlation of the out-of-sample error. The placebo-scale intervals include zero in California and West Germany. The prediction intervals of \citet{cattaneo2021} for the average effect are $[-16.6,3.4]$, $[-2{,}940,-469]$ and $[-1.93,-0.59]$. They are longer than the linear-class intervals and shorter than the curvature intervals at $C_{\mathrm{don}}$, which is the ordering the two classes should produce, since a bound on the post-treatment error calibrated from its pre-treatment distribution restricts the drift less than linearity does and more than a curvature bound set by the most curved placebo does. The breakdown curvature says where between the two a given conclusion stops holding.

The $t$-test of \citet{chernozhukov2021t} with three blocks of the pre-treatment window gives $[-45.2,15.6]$, $[-3{,}164,-568]$ and $[-1.46,-0.08]$ with de-meaned weights, and with simplex weights it excludes zero in California, $[-33.3,-6.8]$, where the trend-balancing interval, the placebo-calibrated interval with the same weights and the prediction interval all include it. Its blocks of six, ten and five years are short relative to the drift in these panels, and the simulations of Section~\ref{sec:endtoend} show that it undercovers when the treated unit's trend differs from the synthetic unit's. The conformal set with cyclic permutations is $[-99,91]$ packs in California, empty in West Germany and $[-3.7,1.0]$ in the Basque Country, which is the behaviour of a permutation test on residuals that drift.

Three cautions bound what these results show. The placebo curvatures are twice the quadratic coefficient of a fitted quadratic, an average curvature estimated from noisy gaps, whereas Assumption~\ref{ass:curv} bounds the supremum of $|q''|$, so the counts of placebos above $C^*$ are descriptive and not a calibration of the class. Assumption~\ref{ass:exch} is a statement about the joint law of the units and cannot be checked from one realisation, and the simulations show that it fails when the treated unit drifts more than every donor. And the noise variance is held fixed across the validation block and the post-treatment years, which Remark~\ref{rem:hetero} relaxes only under a known bound on their ratio.

\section{Discussion}\label{sec:discussion}
The coverage statement separates two commitments. The sensitivity constants restrict how far the untreated comparison may depart from the maintained path, and the Gaussian sampling model determines how noise enters the estimator and the scale. A precise standard error cannot compensate for an invalid counterfactual restriction, and a credible restriction does not by itself supply a reference distribution. The level allowance $L$ deserves attention because a shift in the untreated discrepancy at the intervention date is confounded with a constant treatment effect, so its bound is an identifying restriction rather than a nuisance.

The split, the fitting rule and the anchor rule should be fixed before the validation and post-treatment outcomes are inspected. Searching over splits or anchors for the narrowest interval changes the selection event and falls outside Theorem~\ref{thm:split}, with the one exception that Theorem~\ref{thm:exch} makes explicit, where the simultaneous critical value pays for choosing the anchor within a stated family. The anchor weights and the residual degrees of freedom should be reported with the interval, because they show where its precision comes from.

The guarantees fail in predictable ways. A placebo-calibrated bound, for drift or for curvature, is worth what the exchangeability of the treated unit with its donors is worth, and it fails when the treated unit drifts or bends more than every donor, as in designs C and E2. Persistent disturbances make the Student calibration undercover, and the calibrations of Section~\ref{sec:dependence} restore validity over a stated class at a price in length. The breakdown curvature and the breakdown persistence are the honest response to both problems in a single application. They do not remove the dependence of the conclusion on the restriction, but they state it in units a reader can judge. Extensions to non-Gaussian disturbances by a self-normalised argument on the validation residuals, to several treated units sharing a validation block, and to a placebo-calibrated curvature bound, which would need a pivot for the treated unit's curvature after treatment, each require their own probability argument.

\section{Conclusion}\label{sec:conclusion}
Imperfect synthetic-control fit creates an inference problem whose resolution depends on explicit restrictions on the untreated discrepancy. Under a Gaussian model with time independence, selecting the donor weights and the temporal anchor on a training block yields an exact conditional decomposition into bounded bias and sampling noise, and an orthogonal projection of the validation residuals yields an independent scale. The resulting Student interval has finite-sample coverage uniformly over the class. With the slope left free and the curvature bounded, the breakdown curvature states how much bending of the counterfactual path a conclusion needs, and the donors' placebo curvatures say whether that much is plausible. A drift bound calibrated from the donors is valid under placebo exchangeability at a price of at most $1/(J+1)$. In the classic applications the reunification result survives any curvature the donors display when 1990 is treated as a normal year, and the California result very little. The practical recommendation is to report the estimate with its anchor, its residual degrees of freedom, its breakdown curvature and its breakdown persistence.

\bibliographystyle{plainnat}
\bibliography{references}
\clearpage
\appendix
\section{Proofs}\label{app:proofs}

\begin{proof}[Proof of Proposition~\ref{prop:decomposition}]
Substitution of~\eqref{eq:observed} and~\eqref{eq:discrepancy} into~\eqref{eq:estimator} eliminates $c$, because $\one_n'\lambda=1$, and gives~\eqref{eq:decomposition}. Conditional on $\F$, the weights and nuisance quantities are fixed. Independence across periods and~\eqref{eq:gapvariance} imply~\eqref{eq:normal}. The triangle inequality gives $|\beta|\le L+M|h|+S\norm{\lambda_{\K}}_1$. Equality at the positive endpoint is obtained by taking $\ell=L$, $d=M\operatorname{sgn}(h)$, and $\kappa_t=-S\operatorname{sgn}(\lambda_t)$ on $\K$, with arbitrary permitted signs for zero coordinates. Reversing these signs gives the negative endpoint. These are extrema over the stated rectangular nuisance set.
\end{proof}

\begin{proof}[Proof of Lemma~\ref{lem:student} and Theorem~\ref{thm:split}]
Conditional on $\F$, write $H=I_n-P_\lambda$. Since $H\one_n=Hx_{\Ptwo}=H\lambda=0$,
\[
 e_\lambda=H\kappa+H\eta_{\Ptwo}.
\]
Choose an $n\times\nu_\lambda$ matrix $U$ with orthonormal columns spanning $\col(H)$. Then $UU'=H$ and $U'\eta_{\Ptwo}/\omega\sim N(0,I_{\nu_\lambda})$. Consequently,
\[
 \frac{e_\lambda'e_\lambda}{\omega^2}
 =\norm{\frac{U'\kappa}{\omega}+\frac{U'\eta_{\Ptwo}}{\omega}}^2
 \sim\chi^2_{\nu_\lambda}(\zeta_\lambda).
\]
Furthermore, $\Cov(\lambda'\eta_{\Ptwo},H\eta_{\Ptwo}\mid\F)=\omega^2\lambda'H=0$. Joint Gaussianity gives independence of these two components, and the post-treatment average is independent of both. Adding the fixed vector $H\kappa$ preserves independence of $e_\lambda$ and $\xi(\lambda)$.

For stochastic dominance, a noncentral chi-square variable has the mixture representation $\chi^2_{\nu}(\zeta)\overset d=\chi^2_{\nu+2N_\zeta}$, where $N_\zeta\sim\operatorname{Poisson}(\zeta/2)$. Conditional on $N_\zeta=j$, couple this variable with a central $\chi^2_\nu$ plus an independent $\chi^2_{2j}$ term, interpreted as zero when $j=0$. This proves first-order stochastic dominance.

Let $W_\zeta=\widehat\sigma_\lambda/\sigma(\lambda)$ and let $W_0\sim\sqrt{\chi^2_{\nu_\lambda}/\nu_\lambda}$, independent of $Z\sim N(0,1)$. For $k=t_{\nu_\lambda,1-\alpha/2}$,
\[
 \Prob(|Z|\le kW_\zeta\mid\F)
 \ge\Prob(|Z|\le kW_0)=1-\alpha.
\]
The inequality uses stochastic dominance and monotonicity of $2\Phi(kw)-1$ in $w$. Since $|\widehat\theta-\theta|\le B+|\xi|$, this event implies coverage by~\eqref{eq:honest-ci}. The argument proves~\eqref{eq:conditional-coverage} for every admissible law. Integration over $\F$ proves honesty. The rejection event in~\eqref{eq:test} is the complement of coverage under $\theta=\theta_0$, which proves uniform size control.
\end{proof}

\begin{proof}[Proof of Corollary~\ref{cor:don} and the calibration statement in Remark~\ref{rem:calibration}]
A zero-variance mean-zero Gaussian donor disturbance is identically zero. Donor-only quantities are therefore deterministic functions of the fixed mean paths. They do not change the conditional disturbance law, and Theorem~\ref{thm:split} applies whenever its sensitivity restrictions hold. For probabilistically valid bounds, let $E$ be the training-measurable event on which those restrictions hold. Conditional coverage on $E$ is at least $1-\alpha_0$, so unconditional coverage is at least $(1-\alpha_0)\Prob(E)\ge1-\alpha_0-\gamma$.
\end{proof}

\begin{proof}[Proof of Proposition~\ref{prop:oracle} and equations~\eqref{eq:k-calibration}--\eqref{eq:k-limits}]
For $c>0$, the function $\Prob(|Z+a|\le c)=\Phi(c-a)-\Phi(-c-a)$ is even in $a$ and nonincreasing in $|a|$. Therefore the least-favourable shifts in $[-B/\sigma,B/\sigma]$ are its endpoints. Equation~\eqref{eq:folded} establishes coverage and equality there. Strict increase of this coverage probability in $c$ proves minimality of the half-length for the scalar experiment.

For the Student-scale interval, put $a=\beta/\sigma$ and $r=B/\sigma$. Conditional on $W$, the coverage event is $|Z+a|\le r+kW$, with its minimum at $a=\pm r$. At $a=r$, coverage is
\[
 \E[\Phi(kW)-\Phi(-2r-kW)].
\]
For fixed $r$, this expression is continuous and strictly increasing in $k\ge0$, starts below $1/2$, and converges to one. It therefore has a unique solution for $1-\alpha>1/2$. For fixed $k$, it is nondecreasing in $r$, implying that the solution is nonincreasing in $r$. At $r=0$, the equation is $\Prob(|Z/W|\le k)=1-\alpha$. As $r\to\infty$, dominated convergence gives $\Prob(Z/W\le k)=1-\alpha$. These establish~\eqref{eq:k-limits}. If the scale is noncentral, stochastic dominance preserves the lower coverage bound for a fixed known $r$.
\end{proof}

\begin{proof}[Proof of Proposition~\ref{prop:power}]
Conditional rejection probability equals
\[
 1-\Phi\!\left(c_\alpha-\frac{\Delta+\beta}{\sigma}\right)
 +\Phi\!\left(-c_\alpha-\frac{\Delta+\beta}{\sigma}\right).
\]
It is even and nondecreasing in $(\Delta+\beta)/\sigma$ in absolute value. The smallest possible absolute shift over $\beta\in[-B,B]$ is $d_\Delta$, giving~\eqref{eq:power}. Since $c_\alpha\le B/\sigma+z_{1-\alpha/2}$ by Lemma~\ref{lem:cv},
\[
 d_\Delta-c_\alpha\ge\frac{|\Delta|-2B}{\sigma}-z_{1-\alpha/2}
\]
whenever $|\Delta|\ge B$. The stated separation condition makes the right-hand side diverge and the lower bound on power converge to one.
\end{proof}

\begin{proof}[Proof of Lemma~\ref{lem:cv}]
Let $F(c,b)=\Phi(c-b)-\Phi(-c-b)$. Implicit differentiation of $F(f_\alpha(b),b)=1-\alpha$ yields
\[
 f_\alpha'(b)=\frac{\varphi(f_\alpha(b)-b)-\varphi(f_\alpha(b)+b)}
 {\varphi(f_\alpha(b)-b)+\varphi(f_\alpha(b)+b)}
 =\tanh(bf_\alpha(b)).
\]
Differentiation once more gives the displayed strictly positive second derivative. Since $F(c,b)<\Phi(c-b)$, $f_\alpha(b)-b>z_{1-\alpha}$. Also $f_\alpha'(b)<1$, so $f_\alpha(b)-b$ decreases from $z_{1-\alpha/2}$. Its limit must be $z_{1-\alpha}$ by the defining equation and $f_\alpha(b)+b\to\infty$.

The map $\chi_\alpha(B,\sigma)=\sigma f_\alpha(B/\sigma)$ is the perspective of a convex function and is jointly convex for $\sigma>0$. Differentiation gives~\eqref{eq:chi-derivatives}. In particular,
\[
 g_\alpha(b)>f_\alpha(b)-b>z_{1-\alpha}\quad(b>0),
 \qquad g_\alpha'(b)=-b f_\alpha''(b)\le0,
\]
and $g_\alpha(0)=z_{1-\alpha/2}>z_{1-\alpha}$. Thus $\chi_\alpha$ is strictly increasing in $\sigma$ and nondecreasing in $B$. The bounds on $f_\alpha(b)-b$ give~\eqref{eq:chi-bounds}.
\end{proof}

\begin{proof}[Proof of Theorem~\ref{thm:affine}]
The bias bound is convex in $\lambda$. Moreover,
\[
 \sigma_R(\lambda)=\omega\norm{\begin{pmatrix}m^{-1/2}\\R^{1/2}\lambda\end{pmatrix}}
\]
is strictly convex. Indeed, equality in convexity of the Euclidean norm requires positive collinearity of the two augmented vectors. Their identical nonzero first coordinate forces the proportionality factor to be one, and $R\succ0$ then forces equality of the anchors. Joint convexity of $\chi_\alpha$, its monotonicity in both arguments, and its strict monotonicity in $\sigma$ imply strict convexity of the composition. Equation~\eqref{eq:chi-bounds} and positive definiteness of $R$ imply coercivity. Continuity and closedness of $\Lambda_n$ give existence and uniqueness.

For a fixed value of $h$, the constraints are $A'\lambda=v(\bar x_{\Q}-h)$. Minimization of $\lambda'R\lambda$ under these constraints gives
\[
 \lambda(h)=R^{-1}A(A'R^{-1}A)^{-1}v(\bar x_{\Q}-h),
 \qquad \lambda(h)'R\lambda(h)=q(\bar x_{\Q}-h).
\]
This follows either from the Lagrange equations or by decomposing any feasible vector into $\lambda(h)$ and an $R$-orthogonal constraint-null component. Since $\chi_\alpha$ is strictly increasing in scale, this anchor is optimal at fixed $h$. The map $h\mapsto\lambda(h)$ is affine and injective, so $\phi(h)=F(\lambda(h))$ is strictly convex.

The quadratic $q$ is uniquely minimized at $a_0=-G_{12}/G_{22}$, which equals the expression in~\eqref{eq:q}. Thus $s$ is uniquely minimized at $h_0$. If $h_0\ge0$, every $h<0$ has both a weakly greater bias bound and a strictly greater scale than $h=0$, and every $h>h_0$ is similarly dominated by $h=h_0$. Hence $h^*\in[0,h_0]$. The case $h_0<0$ follows by reversing the sign of $x$. When $M=0$, minimizing $\chi_\alpha(L,\sigma_R)$ is equivalent to minimizing variance, giving~\eqref{eq:mv-anchor}. When $h_0=0$, that anchor minimizes both arguments of $\chi_\alpha$ for every $M$.

Suppose $h_0>0$, $L=0$, and $M>0$. On $[0,h_0]$, set $b=Mh/s(h)$. Then
\[
 \phi'(h)=M f_\alpha'(b)+g_\alpha(b)s'(h).
\]
At $h=0$, $f_\alpha'(0)=0$ and $s'(0)<0$, so $\phi'(0+)<0$. At $h=h_0$, $s'(h_0)=0$ and $b>0$, so $\phi'(h_0)>0$. Strict convexity yields $0<h^*<h_0$. Further,
\[
 \frac{\partial\phi'(h)}{\partial M}
 =f_\alpha'(b)+\frac{h}{s(h)}f_\alpha''(b)\{M-bs'(h)\}\ge0,
\]
because $s'(h)\le0$ on this interval. The unique derivative zero is therefore nonincreasing in $M$. Finally, $M h^*\le\phi(h^*)\le\phi(0)$, where the last quantity is independent of $M$, proving $h^*\to0$.

If $L>0$, $b_0=L/s(0)>0$ and $f_\alpha'(b_0)>0$. By convexity on $[0,h_0]$, its left endpoint is optimal exactly when
\[
 \phi'(0+)=M f_\alpha'(b_0)+g_\alpha(b_0)s'(0)\ge0.
\]
Differentiating~\eqref{eq:sh} gives the formula for $s'(0)$ and hence~\eqref{eq:M-threshold}. The expression for $R=I_n$ follows by inverting $A'A$.
\end{proof}

\begin{proof}[Proof of Proposition~\ref{prop:exclusion}]
Adding $S\norm{\lambda_{\K}}_1$ preserves convexity of the bias bound, so strict convexity and coercivity follow by the preceding argument. The restricted feasible set is a nonempty closed affine set because $\K$ is proper. Convex subgradient optimality therefore gives $v$ and $a$ satisfying~\eqref{eq:restricted-KKT}. At $\lambda^\circ$, the total bias bound and scale are independent of $S$. The chain rule gives
\[
\partial F_S(\lambda^\circ)
 \supseteq g+f_\alpha'(b^\circ)S\,\partial\norm{\lambda_{\K}^\circ}_1,
\]
where the last subdifferential has coordinates in $[-1,1]$ on $\K$ and zero elsewhere, and $g$ is the selected subgradient from~\eqref{eq:restricted-KKT}. Since $b^\circ>0$, $f_\alpha'(b^\circ)>0$. When~\eqref{eq:S-threshold} holds, choose each contaminated-coordinate subgradient to cancel $g_i+a$. All uncontaminated coordinates already satisfy stationarity. Thus $0\in\partial F_S(\lambda^\circ)+a\one_n$, which is sufficient for global optimality on $\Lambda_n$. Strict convexity gives uniqueness.
\end{proof}

\begin{proof}[Proof of Proposition~\ref{prop:soft}]
For a fixed total contaminated weight $q=\sum_{t\in\K}\lambda_t$, equal weights within each group minimize $\lambda'\lambda$ by Cauchy--Schwarz. They also attain the lower bound $\norm{\lambda_{\K}}_1\ge|q|$. Monotonicity of $\chi_\alpha$ therefore reduces the problem to
\[
 \min_{q\in\R}\ \chi_\alpha(S|q|,s_c(q)),\qquad
 s_c(q)=\omega\left\{m^{-1}+\frac{(1-q)^2}{n_0}+\frac{q^2}{k}\right\}^{1/2}.
\]
The scale is minimized at $q=k/n$. Values below zero are dominated by zero, and values above $k/n$ are dominated by $k/n$. At zero, the right derivative is $z_{1-\alpha/2}s_c'(0)<0$. At $k/n$, the derivative is $S f_\alpha'(S(k/n)/s_c(k/n))>0$. Strict convexity gives the interior solution in~\eqref{eq:soft}.

Optimality relative to $q=0$ and~\eqref{eq:chi-bounds} imply $q_S=O(S^{-1})$, so $q_S\to0$. Put $b_S=S q_S/s_c(q_S)$. If $b_S$ stayed bounded away from zero along a subsequence, strict increase of $f_\alpha$ would make its limiting objective exceed $s_c(0)f_\alpha(0)$, contradicting optimality. Thus $b_S\to0$. The first-order condition is
\[
 S f_\alpha'(b_S)+g_\alpha(b_S)s_c'(q_S)=0.
\]
Using $f_\alpha'(b)=z_{1-\alpha/2}b+o(b)$ and $g_\alpha(b)\to z_{1-\alpha/2}$ gives $S b_S\to-s_c'(0)$. Consequently,
\[
 S^2q_S\longrightarrow-s_c(0)s_c'(0)=\frac{\omega^2}{n_0},
\]
as claimed.
\end{proof}

\begin{proof}[Proof of Proposition~\ref{prop:dof}]
The correctly specified mean $Zb_0$ lies in $\col(Z)$, so $r=(I_N-P_Z)v$. Under~\eqref{eq:regressioncondition},
\[
 \E[r'r\mid Z]
 =\operatorname{tr}\!\left((I_N-P_Z)\E[vv'\mid Z]\right)
 =\omega^2\operatorname{tr}(I_N-P_Z)=\omega^2(N-p).
\]
If $p=N$, the residual-maker matrix is zero. Every constrained residual sum of squares is at least the unrestricted minimum. Evaluating the constrained criterion at the true slopes and refitting the unrestricted intercept gives $\sum_{t=1}^N(v_t-\bar v)^2$, whose conditional expectation is $\omega^2(N-1)$. These bounds also apply to an infimum, without requiring existence of a minimizing coefficient vector.
\end{proof}

\begin{proof}[Proof of Proposition~\ref{prop:curv}]
For $x\ge x_0$, Taylor's formula with integral remainder and $q(x_0)=q'(x_0)=0$ give $q(x)=\int_{x_0}^{x}(x-s)q''(s)\,ds=\int K(x,s)q''(s)\,ds$, and for $x<x_0$ the same formula gives $q(x)=\int_{x}^{x_0}(s-x)q''(s)\,ds=\int K(x,s)q''(s)\,ds$. With $\lambda\in\Lambda^0_n$ the slope term $d\,h(\lambda)$ vanishes for every $d\in\R$, and for $\lambda\notin\Lambda^0_n$ it is unbounded, which is why the admissible anchors are the trend-balanced ones. The curvature component of the conditional bias is the linear functional $m^{-1}\sum_{\Q}q(x_t)-\sum_{\Ptwo}\lambda_tq(x_t)=\int k_\lambda(s)q''(s)\,ds$, and $|\int k_\lambda q''|\le C\int|k_\lambda|=C\kappa_C(\lambda)$ by H\"older's inequality, with equality at $q''=C\operatorname{sign}(k_\lambda)$ almost everywhere, which is admissible since the class only requires $q'$ to be Lipschitz. The remaining components are bounded as in Proposition~\ref{prop:decomposition} and the bounds add. The coverage statements use $|\E[\widehat\theta\mid\F]-\theta|\le B$, the law of the disturbances and, for the scale, the fact that the validation residual is a centred Gaussian vector plus an $\F$-measurable shift, here $(I_n-P_\lambda)\{q(x_{\Ptwo})+\kappa\}$, to which the dominance step of the proof of Theorem~\ref{thm:split} and Anderson's inequality apply. For convexity, $k_\lambda(s)$ is affine in $\lambda$ for each $s$, so $\kappa_C$ is an integral of absolute values of affine functions. On a grid containing the observation times $|k_\lambda|$ is convex in $s$ on each knot interval, so the trapezoid rule bounds the integral from above and the discretised bound is conservative. The discretised bound is a finite sum of absolute values of affine functions of $\lambda$, $a(\lambda)$ is a Euclidean norm, and $\Lambda^0_n$ is affine, so the programme is a second-order cone programme with an attained minimum. The finite-$M$ variant adds the convex term $M|h(\lambda)|$ and replaces $\Lambda^0_n$ by $\Lambda_n$.
\end{proof}

\begin{proof}[Proof of Theorem~\ref{thm:exch}]
By Proposition~\ref{prop:decomposition} with $S=0$, $\widehat\theta(\lambda(h))-\theta=\ell+d\,h+\xi(\lambda(h))$ for every $h\in[0,h_U]$, since $h(\lambda(h))=h$, and in particular $\Pi_1=\ell+d\,h_U+\xi(\lambda_U)$. Let $E_1=\{|\Pi_1|\le W\}$ and $E_2=\{\sup_{h\in[0,h_U]}|\xi(\lambda(h))|/\widehat\sigma_{\lambda(h)}\le k^{\mathrm{sim}}_\alpha\}$. On $E_1\cap E_2$,
\[
|d|\,|h_U|\le|\Pi_1|+|\ell|+|\xi(\lambda_U)|\le W+L+k^{\mathrm{sim}}_\alpha\widehat\sigma_{\lambda_U},
\]
so $|d|\le\widehat M$, and for every $h\in[0,h_U]$
\[
|\widehat\theta(\lambda(h))-\theta|\le L+\widehat M|h|+|\xi(\lambda(h))|\le L+\widehat M|h|+k^{\mathrm{sim}}_\alpha\widehat\sigma_{\lambda(h)} .
\]
The inclusion therefore holds simultaneously for all $h$, and so for any random $\widehat h$. Conditionally on $\F$ the anchors $\lambda(h)$ are fixed vectors in $\col([\one_n,x_{\Ptwo}])$, the residual $(I_n-P_{[\one_n,x_{\Ptwo}]})u_{\Ptwo}$ equals $(I_n-P_{[\one_n,x_{\Ptwo}]})\eta_{\Ptwo}$ because $\kappa=0$, and the vector of ratios $\xi(\lambda(h))/\widehat\sigma_{\lambda(h)}$ is a function of $(\eta_{\Ptwo},\bar\eta_{\Q})/\omega$ alone, whose law under Assumption~\ref{ass:prim} does not depend on $\F$, $\omega$ or the discrepancy. Hence $\Prob(E_2^c\mid\F)=\alpha$ by the definition of $k^{\mathrm{sim}}_\alpha$ when the supremum has a continuous law, and at most $\alpha$ otherwise, so $\Prob(E_2^c)\le\alpha$. Assumption~\ref{ass:exch} gives $\Prob(E_1^c)\le1/(J+1)$. The union bound, which uses no independence between $E_1$ and $E_2$, completes the proof.
\end{proof}

\begin{proof}[Proof of Proposition~\ref{prop:dependence}]
Write $\xi=\xi(\lambda)=a'\eta$ with $a=(-\lambda',m^{-1}\one_m')'$, so that $\Var(\xi\mid\F)=\omega^2a'\bar Ra\le\overline\rho\,\omega^2a(\lambda)^2=\overline\rho\,\sigma^2_{\mathrm{ind}}$, where $\sigma_{\mathrm{ind}}=\omega a(\lambda)$ is the scale under independence. Let $H=I_n-P_\lambda$ and $e=H\kappa+H\eta_{\Ptwo}$. The vector $H\eta_{\Ptwo}$ is Gaussian with covariance $\omega^2HR_{\Ptwo}H$, where $R_{\Ptwo}$ is the principal submatrix of $\bar R$ on the validation periods, whose eigenvalues lie in $[\underline\rho,\overline\rho]$ by interlacing. Let $U$ have orthonormal columns spanning $\col(H)$, so that $e'e=\norm{U'e}^2$ with $U'e=U'\kappa+U'\eta_{\Ptwo}$, where $U'\kappa$ is $\F$-measurable and $U'\eta_{\Ptwo}/\omega\sim N(0,U'R_{\Ptwo}U)$ conditionally on $\F$, a nondegenerate Gaussian vector in $\R^{\nu_\lambda}$ since $U'R_{\Ptwo}U\succeq\underline\rho I$. By Anderson's inequality, for a centred Gaussian vector $G$ and any fixed $\mu$, $\Prob(\norm{G+\mu}\le c)\le\Prob(\norm{G}\le c)$ for every $c$, because the ball is symmetric and convex, so $e'e$ first-order stochastically dominates $\norm{U'\eta_{\Ptwo}}^2$. Diagonalising $U'R_{\Ptwo}U=V\diag(\mu_1,\dots,\mu_{\nu_\lambda})V'$ with $\mu_i\in[\underline\rho,\overline\rho]$ by the Rayleigh bound, $\norm{U'\eta_{\Ptwo}}^2/\omega^2=\sum_i\mu_iZ_i^2$ with independent standard normal $Z_i$, which is at least $\underline\rho\,\chi^2_{\nu_\lambda}$ pathwise. Hence $\nu_\lambda\widehat\omega^2_\lambda/\omega^2$ stochastically dominates $\underline\rho\,\chi^2_{\nu_\lambda}$, and with $c^2=\sigma^2_{\mathrm{ind}}\underline\rho\,\chi^2_{\nu_\lambda,\alpha/2}/\nu_\lambda$,
\[
\Prob(\widehat\sigma_\lambda^2<c^2\mid\F)\le\Prob\big(\chi^2_{\nu_\lambda}<\chi^2_{\nu_\lambda,\alpha/2}\big)=\alpha/2 .
\]
Since $|\widehat\theta-\theta|\le B(\lambda)+|\xi|$, non-coverage requires $|\xi|>k^{\mathrm{dep}}_\alpha\widehat\sigma_\lambda$, and
\[
\Prob(|\xi|>k^{\mathrm{dep}}_\alpha\widehat\sigma_\lambda\mid\F)\le\Prob(\widehat\sigma_\lambda<c\mid\F)+\Prob(|\xi|>k^{\mathrm{dep}}_\alpha c\mid\F)
\le\frac{\alpha}{2}+\Prob\Big(|Z|>k^{\mathrm{dep}}_\alpha\frac{c}{\sqrt{\overline\rho}\,\sigma_{\mathrm{ind}}}\Big)=\alpha ,
\]
because $k^{\mathrm{dep}}_\alpha c/(\sqrt{\overline\rho}\sigma_{\mathrm{ind}})=z_{1-\alpha/4}$ by the definition of $k^{\mathrm{dep}}_\alpha$. The argument used no independence between $\xi$ and $\widehat\sigma_\lambda$.
\end{proof}

\section{Simulation designs}\label{app:sim}

Each replication draws fifteen units observed for twenty-two consecutive periods, the first eleven pre-treatment, split into a training block of five and a validation block of six, then one excluded period and ten post-treatment periods. Untreated log outcomes are $Y_{it}(0)=a_i+g_i(x_t-\bar x)+l_if_t+e_{it}$ with levels $a_i\sim U[8.5,10.5]$, growth rates $g_i\sim N(0.025,0.012^2)$, loadings $l_i\sim U[0.5,1.5]$ on a common first-order autoregressive cycle $f_t$ with coefficient 0.6 and innovation standard deviation 0.02, and independent noise $e_{it}\sim N(0,s_i^2)$ with $s_i\sim U[0.01,0.03]$, so that donors are as noisy as the treated unit. The treated unit's level is set 0.8 below the lowest donor's and its effect is $-0.30$ in every post-treatment period. The common cycle violates Assumption~\ref{ass:prim}, since its imbalance $(l_1-w'l_c)f_t$ is neither drift nor independent noise, and it is kept so that the designs match those of \citet{companion} and so that the experiment is not rigged in the construction's favour. The discrepancy in designs A, C, D and B2 is linear in time, and the curvature designs E1 and E2 of Table~\ref{tab:curv} add a quadratic term to design A as described in the table's notes. Design A draws the treated unit's growth like the donors'. Design C sets it 1.5 points a year below the slowest donor's. Design D gives the treated unit's noise a first-order autoregression with coefficient 0.85 and innovation standard deviation 0.05, so that the treated unit is both persistent and noisier than its donors. Design B2 adds an own shock of $-0.15$, $-0.30$, $-0.25$, $-0.20$ and $-0.15$ in pre-treatment periods 7 to 11, which lie in the validation block. Weights are fitted by de-meaned synthetic control. The bound ``$M$ known'' is the realised drift $|g_1-w'g_c|$ of the fitted weights, the bound ``$M$ placebo'' is the largest absolute placebo error $W$ among the fourteen donors with the same split construction divided by $|h_U|$, and ``$S$ known'' is the largest absolute shock in the validation block with the shock window known. The Theorem 3 rows use \eqref{eq:Mhat} with $L=0$ and $k^{\mathrm{sim}}_{0.05}=3.48$, computed from 400,000 draws on a grid of 201 anchors, and the data-chosen anchor minimises the half-length over that grid. The optimal affine anchor and the curvature-optimal anchor are selected with the training-block scale, the root of the pooled variance of the fourteen placebos' leave-one-period-out prediction errors on the training block net of a line, which uses no validation outcome of any unit. The conformal set with cyclic permutations fits de-meaned synthetic control to all periods under each null value on a grid of step 0.05, and the version with a trend model fits simplex weights with an intercept and a linear trend, uses 999 i.i.d.\ permutations and inverts the test by bisection. The $t$-test of \citet{chernozhukov2021t} uses $K=3$ consecutive blocks of $r=\min\{\lfloor T_0/K\rfloor,T_1\}=3$ periods at the start of the pre-treatment window, fits the weights on the other eight periods, and uses $t_{K-1}$ with the factor $(1+Kr/T_1)^{1/2}$. The prediction interval of \citet{cattaneo2021} is computed with scpi for the average effect over the post-treatment periods, simplex weights with an intercept, misspecification-robust in-sample bounds and Gaussian out-of-sample bounds.

The conditional experiment of Table~\ref{tab:simulation} fixes $n=7$ validation periods at $x_t=-7,\dots,-1$, $m=3$ post-treatment periods with $\bar x_{\Q}=1$, the anchor $\lambda=(-0.10,0.05,0.05,0.10,0.20,0.30,0.40)$, so that $h=2.85$ and $\nu=4$, bounds $M=0.05$, $L=0.15$ and unit noise scale, and places the discrepancy at the boundary of its envelope with $d=M$ and $\ell=L$. Contamination of size $S$ enters the third validation period with the sign that maximises the bias. In Panel B the ten disturbances of the validation and post-treatment periods, taken consecutive, follow a stationary first-order autoregression with coefficient $\phi$ and unit variance, $S=0$ and $50{,}000$ replications. Code for both experiments and for Table~\ref{tab:sim} is in the replication package.

\begin{table}[tp]
\centering\scriptsize\setlength{\tabcolsep}{3pt}
\begin{threeparttable}
\caption{Rejection of a zero effect at the nominal 5 percent level}\label{tab:simpower}
\begin{tabular}{@{}p{4.8cm}cccc@{}}
\toprule
 & A & C & D & B2 \\
\midrule
Conventional, $M$ known & 0.99 & 1.00 & 0.87 & 0.16 \\
Conventional, $M=0$ & 1.00 & 1.00 & 0.98 & 0.54 \\
Conventional, $M$ placebo & 0.98 & 1.00 & 0.85 & 0.15 \\
Conventional, placebo scale, $M=0$ & 1.00 & 1.00 & 0.94 & 0.38 \\
Conventional, placebo scale, $M$ known & 0.98 & 1.00 & 0.83 & 0.17 \\
Split weights, validation noise, normal cv & 0.97 & 1.00 & 0.82 & 0.58 \\
Split, uniform, $M$ known$^{\dagger}$ & 0.96 & 1.00 & 0.81 & 0.06 \\
Split, uniform, $M=0$ & 1.00 & 1.00 & 0.96 & 0.35 \\
Split, uniform, $M$ placebo & 0.90 & 1.00 & 0.71 & 0.02 \\
Split, optimal, $M$ placebo & 0.97 & 0.98 & 0.68 & 0.00 \\
Split, trend-balancing$^{\dagger}$ & 0.95 & 0.92 & 0.61 & 0.00 \\
Theorem 3, uniform$^{\ddagger}$ & 0.77 & 1.00 & 0.49 & 0.00 \\
Theorem 3, data-chosen anchor$^{\ddagger}$ & 0.94 & 0.97 & 0.59 & 0.00 \\
Split, uniform, $M,S$ known$^{\dagger}$ &  &  &  & 0.00 \\
Placebo set, split & 0.96 & 1.00 & 0.81 & 0.38 \\
Conformal (CWZ), cyclic & 0.31 & 1.00 & 0.30 & 0.14 \\
Conformal (CWZ), trend model & 0.42 & 0.39 & 0.02 & 0.00 \\
$t$-test (CWZ), $K=3$, simplex & 0.81 & 1.00 & 0.71 & 0.02 \\
$t$-test (CWZ), $K=3$, de-meaned & 0.96 & 1.00 & 0.58 & 0.01 \\
Prediction interval (CFT), average & 0.99 & 0.75 & 0.60 & 0.00 \\
\bottomrule
\end{tabular}
\begin{tablenotes}[flushleft]\footnotesize
\item \emph{Notes:} Share of replications in which the interval excludes zero, true effect $-0.30$. Rows and designs as in Table~\ref{tab:sim}.
\end{tablenotes}
\end{threeparttable}
\end{table}

\section{Additional application results}\label{app:apps}

Table~\ref{tab:abadie-sc} repeats Table~\ref{tab:abadie} with simplex weights fitted to the outcome path without an intercept, as in the original studies but without their covariates.

\begin{table}[tp]
\centering\scriptsize\setlength{\tabcolsep}{3pt}\renewcommand{\arraystretch}{0.95}
\begin{threeparttable}
\caption{Three classic applications, simplex weights without an intercept}\label{tab:abadie-sc}
\begin{tabular}{@{}l lll@{}}
\toprule
 & California & West Germany & Basque Country \\
\midrule
Donors $J$ & 38 & 16 & 16 \\
Weights, training block & 1970--1979 & 1960--1975 & 1955--1961 \\
Anchor, validation block & 1980--1988 & 1976--1990 & 1962--1969 \\
Largest weights & Connecticut 0.33, Utah 0.32 & Netherlands 0.24, New Zealand 0.18 & Madrid 0.72, Galicia 0.28 \\
Placebo curvature $C_{\mathrm{don}}$ & 0.703 & 30.4 & 0.0055 \\
Breakdown $C^*$, optimal anchor & 0.000 (38 of 38 above) & 76.8 (0 of 16 above) & 0.0070 (0 of 16 above) \\
Breakdown $C^*_{\mathrm{tr}}$, trend-balancing anchor & 0.000 (38 of 38 above) & 14.2 (6 of 16 above) & 0.0068 (0 of 16 above) \\
Breakdown persistence $\bar\phi^*$, linear class & 0.0 & $>0.9$ & $>0.9$ \\
\addlinespace
\multicolumn{4}{@{}l}{\emph{Curvature class, $C=C_{\mathrm{don}}$}} \\
\quad Curvature-optimal anchor & $-$2.0 [$-$42.8, 38.7] & $-$3,204 [$-$5,173, $-$1,236] & $-$1.40 [$-$2.50, $-$0.29] \\
\quad Trend-balancing anchor & $-$5.7 [$-$53.4, 41.9] & $-$1,760 [$-$5,069, 1,549] & $-$1.45 [$-$2.66, $-$0.25] \\
\multicolumn{4}{@{}l}{\emph{Linear class ($C=0$)}} \\
\quad Trend-balancing anchor & $-$5.7 [$-$12.8, 1.3] & $-$1,760 [$-$2,161, $-$1,358] & $-$1.45 [$-$1.60, $-$1.30] \\
\quad Trend-balancing anchor, AR(1), $\bar\phi=0.5$ & $-$5.7 [$-$17.3, 5.8] & $-$1,760 [$-$2,443, $-$1,077] & $-$1.45 [$-$1.70, $-$1.20] \\
\quad Theorem 3, data-chosen anchor & $-$5.7 [$-$14.3, 2.8] & $-$1,760 [$-$2,235, $-$1,284] & $-$1.45 [$-$1.64, $-$1.26] \\
\quad Uniform anchor, $M=0$ & $-$17.4 [$-$19.6, $-$15.2] & 279 [113, 444] & $-$0.24 [$-$0.26, $-$0.22] \\
\multicolumn{4}{@{}l}{\emph{Other constructions}} \\
\quad Conventional, in-sample noise & $-$19.4 [$-$20.6, $-$18.2] & $-$1,673 [$-$1,720, $-$1,627] & $-$0.90 [$-$0.94, $-$0.85] \\
\quad Conventional, placebo scale & $-$19.4 [$-$45.2, 6.3] & $-$1,673 [$-$5,208, 1,862] & $-$0.90 [$-$1.87, 0.07] \\
\quad $t$-test (CWZ), $K=3$ & $-$20.1 [$-$33.3, $-$6.8] & $-$1,827 [$-$3,657, 3] & $-$0.76 [$-$1.54, 0.03] \\
\bottomrule
\end{tabular}
\begin{tablenotes}[flushleft]\scriptsize
\item \emph{Notes:} Average effect over the post-treatment years (1989--2000, 1991--2003, 1970--1997), estimate followed by the 95 percent interval, in packs per capita, US dollars per capita and thousand 1986 US dollars per capita. Weights are simplex weights without an intercept, as in the original studies but without their covariates. Curvature rows use Assumption~\ref{ass:curv} with the slope unrestricted. $C_{\mathrm{don}}$ is the largest placebo curvature, twice the quadratic coefficient of a quadratic fitted to a placebo gap over the validation and post-treatment years. The curvature-optimal anchor minimises the Student length over trend-balanced anchors with the training-block scale. $C^*$ and $C^*_{\mathrm{tr}}$ are breakdown curvatures (Definition~\ref{def:breakdown}), with the number of placebos whose curvature exceeds them. $\bar\phi^*$ is the breakdown persistence of the trend-balancing interval on the grid $0,0.1,\dots,0.9$, with 0 meaning that the interval includes zero under independence. The AR(1) row replaces the Student quantile by $k^{\mathrm{num}}_{0.05}$ of \eqref{eq:k-num} over $\phi\in[0,0.5]$, which assumes that the dependence does not reach back into the training block. The Theorem 3 row uses the bound \eqref{eq:Mhat} with $k^{\mathrm{sim}}_{0.05}$, guarantee $1-0.05-1/(J+1)$. Conventional rows fit the weights on all pre-treatment years, anchor uniformly on them and use the normal critical value with the in-sample residual scale or the root mean square of the full-window placebo estimates. The $t$-test of \citet{chernozhukov2021t} uses three blocks of the pre-treatment window. The prediction interval and the conformal set do not depend on the weighting rule and are in Table~\ref{tab:abadie}.
\end{tablenotes}
\end{threeparttable}
\end{table}

\end{document}